\documentclass[aps,prx,onecolumn,superscriptaddress,showpacs,notitlepage]{revtex4-1}

\usepackage[latin1]{inputenc}
\usepackage[english]{babel}
\usepackage{graphicx}
\usepackage{psfrag}
\usepackage{amssymb}
\usepackage{amsmath}
\usepackage{amsxtra}
\usepackage{amstext}
\usepackage{amscd}
\usepackage{eucal}
\usepackage{color}
\usepackage{bm}
\usepackage{hyperref}
\usepackage{mathtools}
\input{epsf}
\begin{document}
\renewcommand{\figurename}{Figure}

\title{Topologically Protected Loop Flows in High Voltage AC Power Grids}

\author{T.~Coletta}
 \affiliation{School of Engineering, University of Applied Sciences of Western Switzerland, CH-1950 Sion, Switzerland} 

\author{R.~Delabays}
 \affiliation{School of Engineering, University of Applied Sciences of Western Switzerland, CH-1950 Sion, Switzerland} 
 \affiliation{Section de Math\'ematiques, Universit\'e de Gen\`eve, CH-1211 Gen\`eve, Switzerland}
\author{I.~Adagideli}
 \affiliation{Faculty of Engineering and Natural Sciences, Sabanci University, Orhanli-Tuzla, Istanbul, Turkey} 
\author{Ph.~Jacquod}
 \affiliation{School of Engineering, University of Applied Sciences of Western Switzerland, CH-1950 Sion, Switzerland} 
\date{\today}

\begin{abstract}
Geographical features such as mountain ranges or big lakes and inland seas
often result in large closed loops in high voltage AC power grids. 
Sizable circulating power flows
have been recorded around such loops, which 
take up transmission line capacity and dissipate but do not deliver electric power.
Power flows in high voltage AC transmission grids 
are dominantly governed by voltage angle differences between connected buses,
much in the same way as Josephson currents depend on phase differences between
tunnel-coupled superconductors.
From this previously overlooked similarity 
we argue here that circulating power flows in AC power grids are analogous to 
supercurrents flowing in superconducting rings and in rings of Josephson junctions. 
We investigate how circulating power flows can be created and how they behave
in the presence of ohmic dissipation. 
We show how changing operating conditions may generate them, 
how significantly more power is ohmically 
dissipated in their presence and
how they are topologically protected, even in the 
presence of dissipation, so that they persist when operating conditions are returned to their
original values. We identify three mechanisms for creating 
circulating power flows, (i) by loss of stability of the equilibrium state 
carrying no circulating loop flow, (ii) by tripping of a line traversing a large loop
in the network and (iii) by reclosing a loop that tripped or was open 
earlier. 
Because voltage angles are uniquely defined, 
circulating power flows can take on only discrete values, much in the same way as 
circulation around vortices is quantized in superfluids. 
\end{abstract}

\maketitle

\section{Introduction}

Power grids are networks of electrical lines whose purpose is to deliver electric
power from producers to consumers. The ensuing power flows do not usually follow specified 
paths, instead they divide among all possible paths following Kirchhoff's laws. 
Circulating loop flows around closed, geographically constrained
loops have been observed in the North American 
high voltage power grid, sometimes reaching as much as 1GW~\cite{Cas98,Ler03},
delivering no power but dissipating it ohmically.
To try and prevent them, grid operators have issued new market regulations and 
recommended integrating phase angle regulating transformers into the grid~\cite{Whi08}. 
The network conditions under which
circulating power flows emerge, why they are so robust,
how much power they dissipate and whether specific network topologies, if any, could prevent them
in the first place 
are issues of paramount importance which have not been addressed
to date.  
At a conceptual level, the definition of circulating loop flows is furthermore ambiguous, being
arbitrarily based on 
an ill-defined separation of power flows into direct, parallel-path and circulating loop flows.
Our goal
in this manuscript is to understand better the nature of these circulating power 
flows. 

A deep and unexpected 
analogy between high voltage electric power transmission and macroscopic quantum
states such as superfluids and superconductors has been overlooked so far. 
The operational state of an AC power grid is determined by the complex voltage 
at each bus, $V_l = |V_l| \exp[i \theta_l]$
which must be single-valued. Summing over voltage angle differences around any loop in 
the network
must therefore give an integer multiple of $2 \pi$. 
This defines topological winding numbers $q_\alpha$,
\begin{eqnarray}\label{windingNumber}
 q_\alpha &=& \left(2\pi\right)^{-1}\sum_{l=1}^{n_\alpha} |\theta_{l+1} - \theta_l| \, \in \, \mathbb{Z}\, ,
\end{eqnarray}
where the sum runs over all $n_\alpha$ nodes around any (the $\alpha^{\rm th}$) 
loop in the network, 
$ |\theta_{l+1} - \theta_l| $ gives the angle difference along the $l^{\rm th}$ line in this loop,
counted modulo $2 \pi$, and node indices are taken modulo $n_\alpha$, i.e. 
$n_\alpha+1 \rightarrow 1$. The topological meaning of $q_\alpha$ is obvious, as it counts the 
number of times the complex voltage winds around the origin in the complex plane as one goes
around the $\alpha^{\rm th}$ loop. 
The condition $q_\alpha \in \mathbb{Z}$ is the same as
the condition that leads to 
quantization of circulation around superfluid vortices~\cite{Ons49,Fey55} or to flux
quantization through a superconducting ring~\cite{Bye61}. 

The analogy with superconductivity
is complete under the {\it lossless line approximation}, where AC transmission lines
are assumed purely susceptive (see Supplemental Material)~\cite{Ber00}. Then, the active 
power flowing between two nodes $l$ and $m$
is given by
$P_{lm} = B_{lm}  |V_l| |V_m| \sin(\theta_l-\theta_m)$, with the
elements $B_{lm}$ of the susceptance matrix. This is the DC Josephson current that would flow
between two superconductors with phases $\theta_l$ and $\theta_m$,
coupled by a short tunnel junction of transparency 
$T_{lm} = \hbar B_{lm} |V_l| |V_m| /8 e$~\cite{Jos62}.
It has been shown within the lossless line approximation that, in complex networks,
various solutions to the power flow problem 
exist, which differ only by circulating loop currents~\cite{Dor13,Del16}.
Neglecting ohmic dissipation, 
it is therefore expected, and has been reported in simple networks~\cite{Del16,Tay12,Meh14}, 
that AC power grids may carry circulating
loop flows. Topological winding numbers, Eq.~(\ref{windingNumber}), 
lead to the discretization of circulating loop flows, much in the same way as superfluid circulation 
is quantized around a vortex~\cite{Ons49,Fey55}. Therefore we refer to circulating loop flows
as {\it vortex flows} from now on. 
The existence of integer 
winding numbers has the important consequence that 
the vortex flows are topologically protected, much in the same way as 
persistent currents in superconducting loops~\cite{Tin75}. The integer $q_\alpha$
in Eq.~(\ref{windingNumber}) measures the number of times 
the complex voltage $V_l = |V_l| \exp[i \theta_l]$ rotates in the complex plane as one goes around 
a loop in the network. Changing a vortex flow requires to change the number $q_\alpha$ of 
such rotations, i.e. to
untwist $V_l$, which cannot be done smoothly without driving $|V_l| \rightarrow 0$ somewhere. 
It is thus hard to get rid of a vortex flow without topological changes in operating 
electrical networks.

AC power grids however differ from superfluids and superconducting systems 
in at least two significant ways in that (i) whereas vortices 
are generated by external magnetic fields (in a superconductor) 
or sample rotation (in a superfluid), how to create vortex flows
in AC power grids is not quite understood, 
and (ii) superfluids and superconductors are nondissipative quantum fluids, 
whereas AC power lines dissipate ohmically part of the power
they transmit.
The lossless line approximation 
is in fact only partially justified in very high voltage AC power grids, 
where lines have a conductance that is at least ten times
smaller than their susceptance~\cite{Ber00}. 
Still, ohmic losses typically reach 
5--10 \% of the total transported power. It is therefore important to find out
whether the above analogy between high voltage AC power grids and macroscopic quantum
states is at all physically relevant. 
The two main purposes of this
manuscript are therefore (i) to investigate how vortex flows can be created in electric power grids
and (ii) to investigate how resilient they are to the presence of ohmic dissipation. 
Our investigations of creation
mechanisms amplify on the work of Janssens and Kamagate~\cite{Jan03} who succinctly discussed
one of the three mechanisms we identify below. Investigating basins of stability for different 
solutions via the Lyapunov function allows us furthermore to shed analytical light on the
line reclosing mechanism they proposed~\cite{Jan03} and give precise bounds on 
when vortex flows are created in this way. We furthermore find that vortex flows are 
resilient to reasonable amounts of ohmic dissipation typical of high voltage power grids and that
vortex-carrying operating states ohmically dissipate significantly more electric power than 
vortex-free states.

Loop flows in  electric power grids have been investigated before 
in a number of theoretical and numerical
works. We list some of the most important published works we know about. 
Korsak investigated a simple network where 
different, linearly stable solutions exist that differ by some circulating loop current~\cite{Kor72}.
Tavora and Smith related the existence of different stable fixed points of the power flow problem 
to the presence of integer winding numbers~\cite{Tav72b}, reflecting the $2 \pi$-periodicity
of the complex voltage around any loop in the network and its single-valuedness. 
The characterization of circulating loop flows with topological winding numbers has been 
pushed further by Janssens and Kamagate~\cite{Jan03}, who also investigated how to 
generate such loop flows and found one of the three creation mechanisms we discuss below. 
More recently, Refs.~\cite{Dor13,Del16} showed that within the lossless line approximation, 
different power flow solutions must be related to one another by circulating loop flows. 

While vortex flows and the analogy we just pointed out
between macroscopic quantum states and high voltage AC power grids are
intellectually interesting in their own right, we stress that they are physically and technically 
relevant. Circulating loop flows in the GW range 
have been observed in power grids~\cite{Cas98,Ler03}, which delivered no power but consumed
it ohmically. Furthemore it can be expected that with the changes in operational conditions 
of the power grid brought about by the energy transition -- substituting delocalized 
productions with smaller primary power reserve 
for large power plants -- such circulating loop flows may occur more frequently. 
We stress, however, that, while power flow solutions are similar to vortex-carrying quantum mechanical
states, there is no quantumness in the power flow problem and that high voltage AC power grids
are not superconducting. 

The paper is organized as follows. In Section~\ref{section2} we argue that 
different solutions to the power
flow problem in meshed networks are related to one another via vortex flows. 
Sections~\ref{section3}, \ref{section4} and \ref{section5} next discuss sequentially the three mechanisms
we identified for generating vortex flows. Section~\ref{section5} in particular investigates
vortex flow creation via reclosing of a line from a basin of attraction point of view, which allows
to quantitatively understand how vortex flows are born.
Section~\ref{section6} makes the case that vortex flows are robust against the relatively
modest amount of ohmic dissipation present in high voltage power networks. 
Having gained much understanding of vortex flows in simple models in these early Sections, we next
discuss vortex creation with and without dissipation on a complex network with the topology
of the UK grid. This is done in Section~\ref{section7}. Conclusions and future perspectives
are briefly discussed in Section~\ref{section8}. 

\section{Circulating Loop Flows in Meshed Networks}\label{section2}

We start with the lossless line approximation and neglect voltage variations, $|V_l| = V_0$, $\forall l$.
Active power flows are then governed by the following set of equations (see Supplemental Material)
\begin{eqnarray}\label{eq:pflow_jj_active}
P_l &=& \sum_{m}
\tilde{B}_{lm} \sin(\theta_l - \theta_m)\, .
\end{eqnarray}
Here, $P_l$ is the active power injected ($P_l>0$) or consumed ($P_l<0$) at node $l$,
$\theta_l$ is the complex voltage angle and 
$\tilde{B}_{lm} = B_{lm}  V_0^2$.
At this level, one has an exact balance between production and consumption,
$\sum_l P_l = 0$.
The theorem of 
Refs.~\cite{Dor13,Del16} states that different solutions to Eq.~(\ref{eq:pflow_jj_active})
on a meshed network differ only by  
circulating flows around loops in the network. We go beyond the lossless line approximation
to see how much validity this theorem keeps in the presence of ohmic losses.
Being interested in
high voltage grids we neglect voltage fluctuations all through this manuscript, as 
they correspond to few percents of the rated voltage $V_0$. Ohmic losses are
introduced by rewriting 
Eq.~(\ref{eq:pflow_jj_active}) as
\begin{eqnarray}\label{eq:pflow_jj_actived}
P_l &=& \sum_{m} \left (
\tilde{B}_{lm} \sin(\theta_l - \theta_m) + \tilde{G}_{lm} [1- \cos(\theta_l - \theta_m) ] \right) \, ,
\end{eqnarray}
where $\tilde{G}_{lm} = G_{lm}  V_0^2$ with elements 
$G_{lm}$ of the conductance matrix.
Because of ohmic dissipation, total production now exceeds total consumption, 
\begin{eqnarray}\label{eq:balance_diss}
\Delta P = \sum_l P_l &=& \sum_{l,m} \tilde{G}_{lm} \, \left[1 - \cos(\theta_l - \theta_m) \right ] > 0\,.
\end{eqnarray}
Eq.~(\ref{eq:balance_diss}) implies in particular that different solutions to Eq.~(\ref{eq:pflow_jj_actived}) 
dissipate different amounts $\Delta P$ of active power and therefore require different power injections
$\{P_l\} \rightarrow \{P_l  + \delta P_l\}$ to compensate for ohmic losses. The set
$\{ \delta P_l\}$ is not uniquely defined, nevertheless, choices exist for which 
the theorem of Refs.~\cite{Dor13,Del16} remains valid (see Supplemental Material). This  
is in agreement
with Baillieul and Byrnes~\cite{Bai82} who stated that "models for lossless power
networks provide valuable insight and understanding for systems with small transfer conductances"
and further suggests 
that circulating loop flows are robust against a moderate amount of ohmic dissipation.
Below we numerically confirm this conjecture. 
Earlier works however suggested that the number of 
solutions decreases at fixed susceptance when the conductance increases~\cite{Ska80,Bai82},
so that it is expected that circulating loop flows are eventually suppressed when the conductance 
exceeds some network-dependent threshold. 

Focusing next on the operational conditions under which circulating
power flows occur, we find three different mechanisms for creating 
them, (i) by loss of stability of the 
solution carrying no circulating loop flow, (ii) by tripping of a line traversing a large loop
in the network and (iii) by reclosing a loop that tripped or was open 
earlier~\cite{Jan03,Col16}.
We discuss
these three mechanisms sequentially, first without ohmic dissipation.

\begin{figure}
 \begin{center}
  \includegraphics[width=425px]{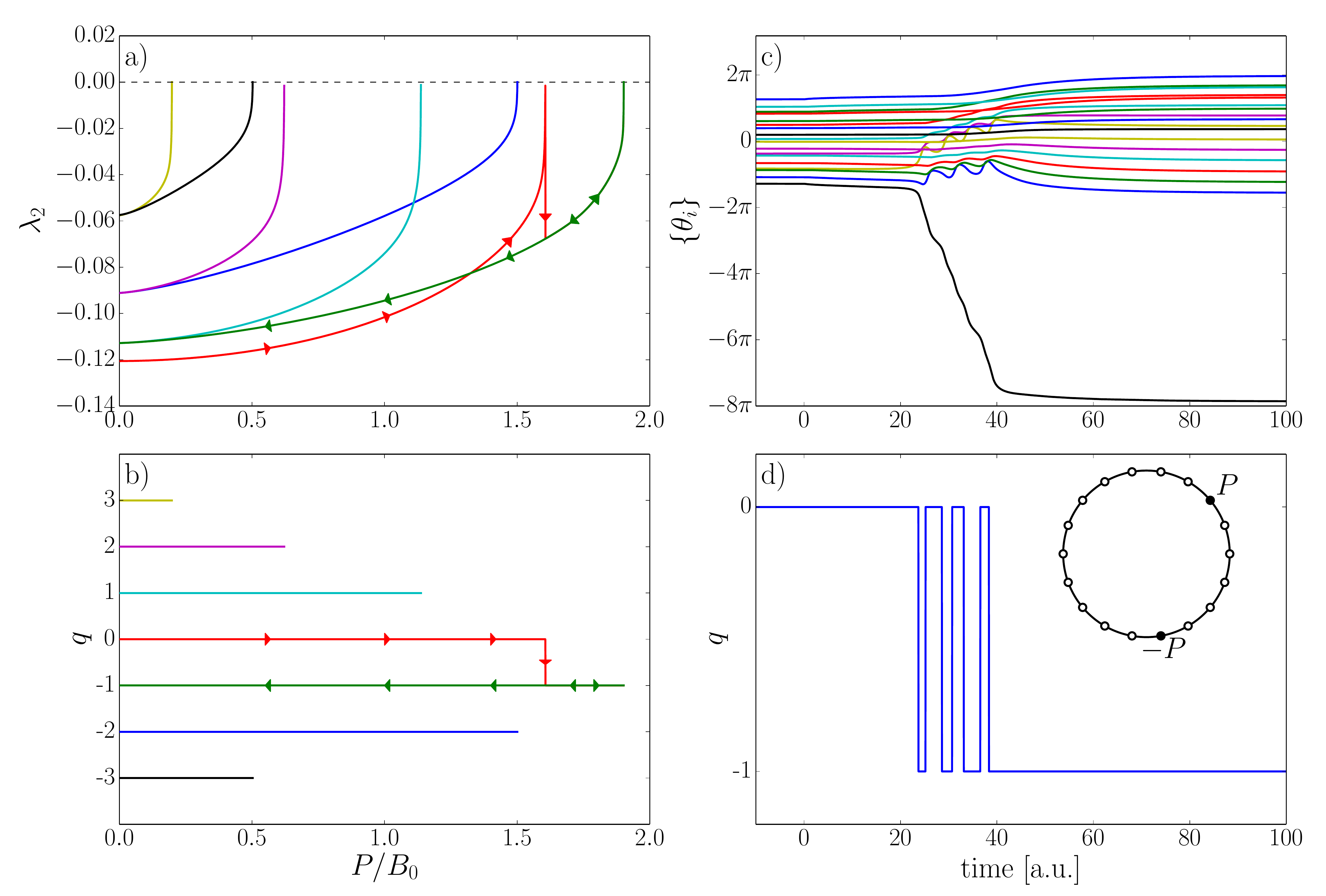}
  \caption{\footnotesize Vortex formation for a ring with $n=18$ nodes. Locations of the
  main power injection and consumption are indicated by the black nodes in the inset of panel d), 
  while all white nodes have small random injections and consumptions summing to zero, to make the 
  model not too specific (see Supplemental Material).
  Panels a) and b): stability diagram for solutions with 
  different winding numbers $q$, showing a) the Lyapunov exponent $\lambda_2$ and b) the range of  
  stability of the solution. Each spike with $\lambda_2 \rightarrow 0$
  in panel a) indicates the  loss of stability of a solution.  
  The red arrows in a) and b) indicate jumps from the $q=0$ 
  solution to that with $q=-1$
  as $P$ increases. The path is not retraced, however, as one
  reverses $P$ back to $P=0$ (green arrows). Panels c) and d): dynamical phase slip as 
  $P/B_0=1.575 \rightarrow 1.65$ at $t=0$. The $q=0$ solution
  loses its stability and Eq.~(\ref{Swing}) induces a transient behavior where
  the angle on the consumption site rotates until 
  one reaches the $q=-1$ solution.}
  \label{fig:Vortex formation by loss of stability}
 \end{center}
\end{figure}

\section{Creating Vortex Flows via Dynamical Phase Slip}\label{section3}

Consider first a single-ring, lossless network as illustrated in 
Fig.~\ref{fig:Vortex formation by loss of stability}d.
The power flow around the ring is governed by Eq.~(\ref{eq:pflow_jj_active}) with
$\tilde{B}_{lm} = B_0$.
From Ref.~\cite{Del16}, the system carries at most nine solutions differing by loop flows, when
$B_0 \rightarrow \infty$. Fig.~\ref{fig:Vortex formation by loss of stability}b shows seven of these solutions, each characterized by a winding number $q=-3, -2, ... 3$.
The stability of each solution is determined by the swing equations~\cite{Ber00}, which
govern the dynamics of the voltage angles $\theta_l$ under changes in operating conditions.
In this work we neglect inertia terms in swing equations, since their presence affects neither
the nature of the stationary states, nor at which parameter values they 
become unstable~\cite{Man14,Col16}.
In a frame rotating at the grid frequency of 50 or 60 Hz, the swing equations 
read (see Supplemental Material)~\cite{Ber81,Pai89,Ber00,Bac13}
\begin{equation}\label{Swing}
\dot{\theta_l}=P_l-\sum_{m=1}^n  B_0 \,  \sin(\theta_l - \theta_m)   \, , \qquad   l=1,\ldots n \, .
\end{equation}
Stationary solutions to Eq.~(\ref{Swing}) are obviously solutions to Eq.~(\ref{eq:pflow_jj_active})
and their linear stability depends on the 
stability matrix $M$ obtained after linearizing Eq.~(\ref{Swing}) (see Supplemental Material)~\cite{Pec98}. 
A solution
of Eq.~(\ref{eq:pflow_jj_active}) is stable if $M$ is negative semidefinite. 
We therefore study the stability of each solution via the largest nonvanishing
eigenvalue $\lambda_2$ of $M$. 
Fig.~\ref{fig:Vortex formation by loss of stability}a shows, together 
with Fig.~\ref{fig:Vortex formation by loss of stability}b how solutions disappear as they lose their
stability, $\lambda_2 \rightarrow 0$. 
The solution with $q=0$ has the smallest $\lambda_2$
at small $P$. Remarkably enough, the $q=0$ solution loses its stability at  $P/B_0\approx1.6$,
before the $q=-1$ solution, which remains stable until $P/B_0\approx1.85$. Starting from the $q=0$ solution
and increasing $P$ beyond 1.6$B_0$, we observe a loss of stability followed by a short transient after
which the operating state has been transferred to the $q=-1$ state. This transient is illustrated in 
Fig.~\ref{fig:Vortex formation by loss of stability}c and d, which show that mostly one voltage angle,
corresponding to the consumer node rotates 
while all other angles move very little (a movie of this transient can be found in the Supplemental Material). 
The rotation of this angle changes $q$ which oscillates between $q=0$ and
$q=-1$, eventually stabilizing at $q=-1$.

Reducing next  $P$ starting from the $q=-1$ solution at $P/B_0 > 1.6$, 
one remains on the $q=-1$ solution. This
hysteretic behavior is indicated by arrows in Fig.~\ref{fig:Vortex formation by loss of stability}a and b and illustrates the topological
protection brought about by the integer winding number $q$. We have found that this
behavior is generic for single-ring networks (see Supplemental Material).
The process by which the winding number changes 
is similar to {\it quantum phase slips}
in small rings of Josephson junctions~\cite{Mat02}. 
To emphasize this similarity, while 
stressing the different physical ingredients at work, we call {\it dynamical phase slip} this first
mechanism of creation of circulating loop flows.

\begin{figure}
\centering
\includegraphics[width=250px]{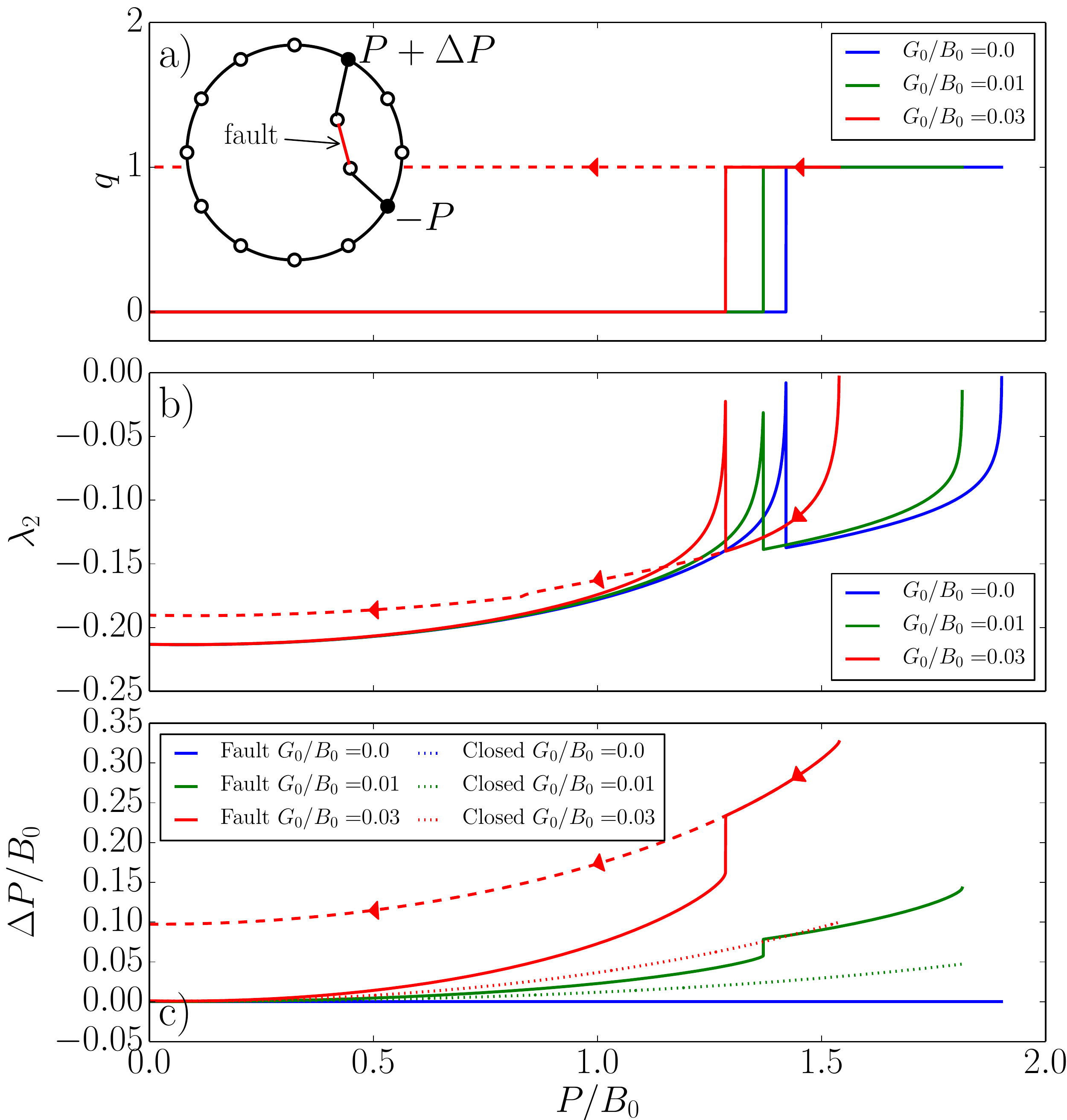}
\caption{\footnotesize Vortex formation by line tripping in a double loop network with $n=14$ nodes
and constant line susceptance $B_0$. Locations of the
  main power injection and consumption are indicated by the black nodes in the inset of panel a), 
  while all white nodes have small random injections and consumptions summing to zero, to make the 
  model not too specific. 
  The initial state corresponds to a $q=0$ state of the double-loop 
  system. The line indicated in red in the inset of panel a) is then tripped, generating a transient to a new
  operating state.
  Panels a) and b) show the winding number and  the
 Lyapunov exponent of the new stationary state, for the lossless case (blue line)
and for dissipative cases with conductances equal to $1\%$ (green line) and $3\%$ (red line)
of the susceptance. The ohmic losses incurred before (dotted lines) and after (solid lines) 
line tripping are compared in panel c). Arrows on all panels indicate that
once created, a vortex flow does not disappear when reducing $P/B_0$.}
\label{fig:DoubleLoop}
\end{figure}

\section{The Line Tripping Mechanism}\label{section4}

We next investigate the second mechanism for vortex flow generation, by tripping
of a line. High voltage AC 
power grids have a meshed structure, where multiple paths connect production and
consumption centers. This ensures that a single line failure does not preclude the
supply of electric power. Consider 
the model shown in the inset of Fig.~\ref{fig:DoubleLoop}a, where a producer is connected to 
a consumer via three different paths.
Assume then that a line on  the middle path trips.
The power initially transmitted via that path is redistributed and if $P$
is relatively large, the angle differences $\Delta_{\rm L,R}$ 
on each remaining path increase significantly. When one 
of these two paths, say the left one, goes through 
many more lines than the other one, $N_{\rm L} \gg N_{\rm R}$, 
it is then possible that $N_{\rm L} \Delta_{\rm L} -  N_{\rm R}\Delta_{\rm R} = 2 \pi q$ with $q > 0$,
even if the system carried no vorticity initially. This simple example shows how one line tripping
in an asymmetrical double-loop system can generate a vortex flow.

Fig.~\ref{fig:DoubleLoop} illustrates how a $q\ne 0$ state emerges from a $q=0$ state
after a line tripping. The initial state is a stationary state of 
the double-loop system with zero winding number on both loops. The red line in the 
inset of Fig.~\ref{fig:DoubleLoop}a is then cut, which induces a transient 
driving the system to a stationary state of the resulting single-loop
system. Fig.~\ref{fig:DoubleLoop}a shows the obtained winding number. One sees that for small
$P$, the final state has $q=0$, while for larger $P$, a $q=1$ state is reached. We have found that
this behavior is generic of sufficiently asymmetric double-loop systems.
We discuss cases with ohmic dissipation below.

\begin{figure}
\centering
\includegraphics[width=250px]{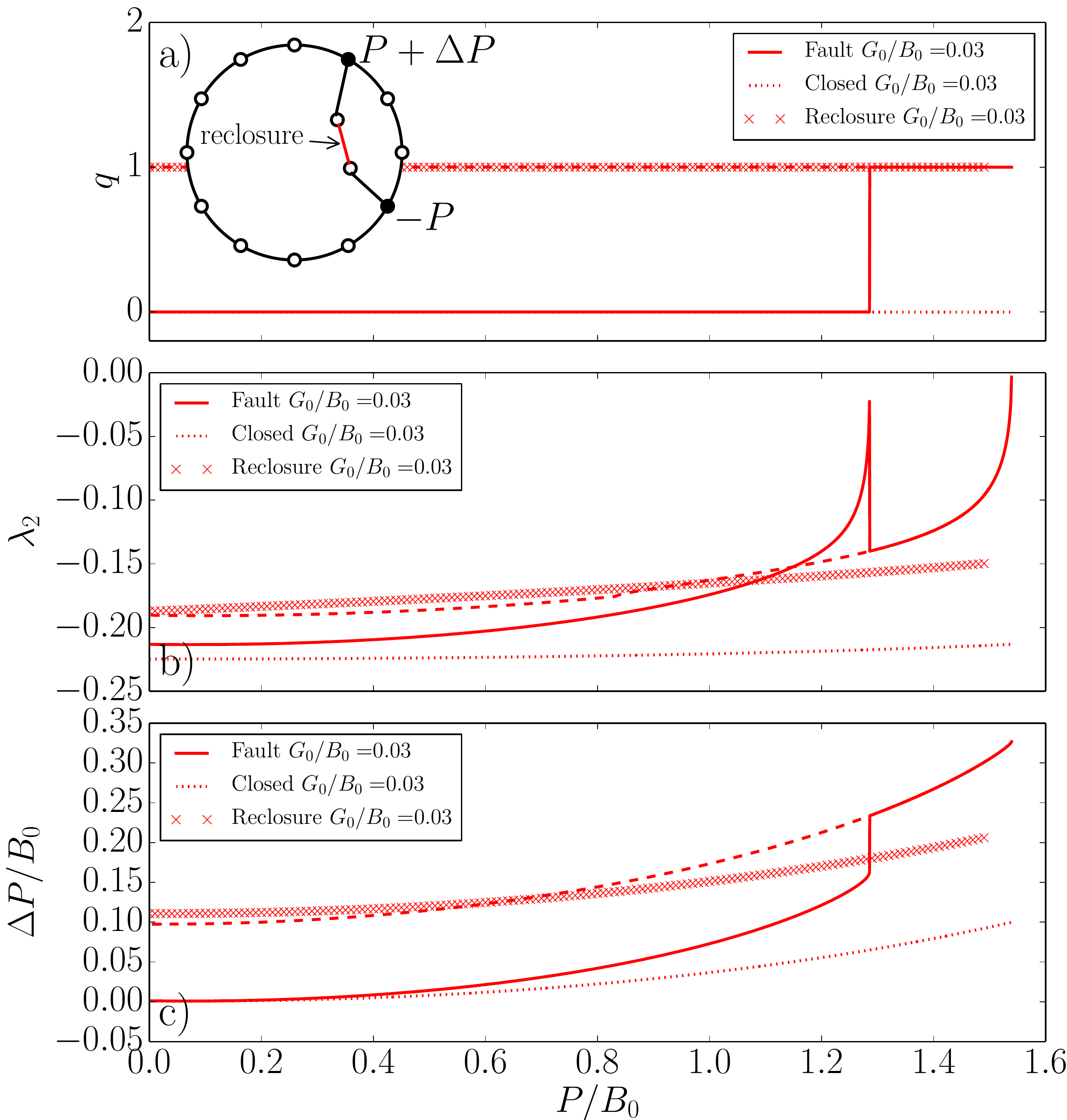}
\caption{\footnotesize Resilience to line reclosing 
of the vortex flow created by the line tripping mechanism
in Fig.~\ref{fig:DoubleLoop}. The solid and dashed red lines are the same as in 
Fig.~\ref{fig:DoubleLoop} and correspond to a conductance equal to $3\%$ of the susceptance. 
The initial state before reclosing is prepared by line tripping at $P/B_0 =1.55$, followed by
reduction of $P/B_0$ (arrows in  Fig.~\ref{fig:DoubleLoop}). It has 
$q=1$ and lies on the dashed red line. The crosses indicate the state obtained 
after line reclosing at a given value of $P/B_0$. 
The vortex flow survives line reclosing and is located in the larger, left
loop.}
\label{fig:DoubleLoopReclosing}
\end{figure}

One may wonder what is the fate of the $q=1$ state created by line tripping 
when the line is reclosed. Line reclosing is a topological change 
that has the potential to induce integer changes in the winding number $q$ so that a vortex-free state
with $q=0$ can be expected after line reclosing. We show in 
Fig.~\ref{fig:DoubleLoopReclosing} that in the present case, line reclosing does not change
the winding number and that the vortex flow persists, and that the winding number
remains the same, $q=1$. Inspecting the angles in the final state, we find that the vortex in the final
state is supported by the larger, left loop. 

\section{Emergence of Vortex Flows from Line Reclosing}\label{section5}

We have just showed that line tripping can lead to a vortex flow that is robust against
the reclosing of the tripped line. In this paragraph, we 
finally consider vortex flow creation via 
reclosing of a line. We consider again the single-loop model 
sketched in the inset of Fig.~\ref{fig:Vortex formation by loss of stability}d. We 
consider here the case with only one consumer and one producer
with some produced
(consumed) power $P$ ($-P$) but checked that our conclusions remain the same and 
that vortex formation proceeds similarly in more complicated single-loop models (see Supplemental Material). 
We start from the closed-loop system with an operating state with $q=0$. 
The power $P$ is transferred from producer to consumer both clockwise (via the right path) and 
counterclockwise (left). One line along 
the right path then trips, which forces $P$ to be transmitted exclusively along the left path. 
The voltage angle difference 
$\Delta_{\rm L}$ between any two nodes 
along the left path increases, while angle differences along the right path 
vanish, $\Delta_{\rm R}=0$. This gives
an angle difference $\Delta_0=N_{\rm L} \Delta_{\rm L}$ 
between the two nodes on each side of the tripped line. 
Upon reclosing that line, a current flows through it whose initial direction depends
on $\Delta_0$, but which eventually relaxes following a dynamical process determined
by Eq.~(\ref{Swing}). The creation of a vortex flow can however be 
understood
without investigating  the voltage angle dynamics, by instead adopting an approach
based on the Lyapunov function~\cite{Pai89,Ber00}. The latter determines the basin of attraction, in 
voltage-angle space, for the various solutions to the power flow problem~\cite{Wil06,Men13}. 
The Lyapunov function corresponding to Eq.~(\ref{Swing}) reads~\cite{Hem93}
\begin{eqnarray}\label{eq:lyapfunction}
{\cal V}(\{ \theta_i \}) &=& -\sum_l P_l \theta_l -\sum_{\langle l,m \rangle} 
B_0\, \cos(\theta_l - \theta_m)  \, ,
\end{eqnarray}
where the second sum runs over connected nodes only.
Minima of ${\cal V}$ have $\nabla_\theta {\cal V} = 0$, and thus
correspond to stationary power flow solutions. 
The Lyapunov function can be rewritten as a function of angle differences 
$\Delta_l\coloneqq\theta_l-\theta_{l+1} \in [-\pi,\pi]$ as
\begin{align}
 \mathcal{V}(\{\Delta_i\})&=-\sum_l P_l^*\Delta_l-\sum_l B_0\cos(\Delta_l)\, ,
\end{align}
where $P_l^*\coloneqq\sum_{j=1}^l P_j$, for $l=1,...,n$.
The single-loop model we consider here can be split in 
two paths (left and right) from producer to consumer. We have
\begin{align}
 P_l^*&=\left\{
 \begin{array}{cl}
  P&\text{if the line from }l\text{ to }l+1\text{ is on the left path,}\\
  0&\text{if the line from }l\text{ to }l+1\text{ is on the right path.}
 \end{array}
\right.
\end{align}
Furthermore, any solution of the nondissipative power flow equations has the same angle differences, 
$\Delta_{\rm L}=\theta_{l}-\theta_{l+1}$, along each line on the left path and $\Delta_{\rm R}=\theta_{l+1}-\theta_{l}$ 
on the right path. 

Going around the loop, the voltage phases must be well defined. Therefore, just before line
reclosing the phase difference $\Delta_0$ between the two ends of the 
tripped line can be written as a function of $\Delta_{\rm L}$ and $\Delta_{\rm R}$, 
$\Delta_0=N_{\rm L}\Delta_{\rm L}-(N_{\rm R}-1)\Delta_{\rm R}-2\pi q$, where $N_{\rm L}>N_{\rm R}\geq2$ are the number of edges 
on the left and right paths. 
Then we can project the Lyapunov function on the $(\Delta_{\rm L},\Delta_{\rm R})$-plane, 
\begin{align}
 {\cal V}(\Delta_{\rm L},\Delta_{\rm R})&=-N_{\rm L}P\Delta_{\rm L}-N_{\rm L}B_0\cos\Delta_{\rm L}-(N_{\rm R}-1)B_0\cos\Delta_{\rm R}-B_0\cos(N_{\rm L}\Delta_{\rm L}-(N_{\rm R}-1)\Delta_{\rm R})\, .\label{si_eq:lyapfunctionplane}
\end{align}

\begin{figure}
 \begin{center}
  \includegraphics[width=425px]{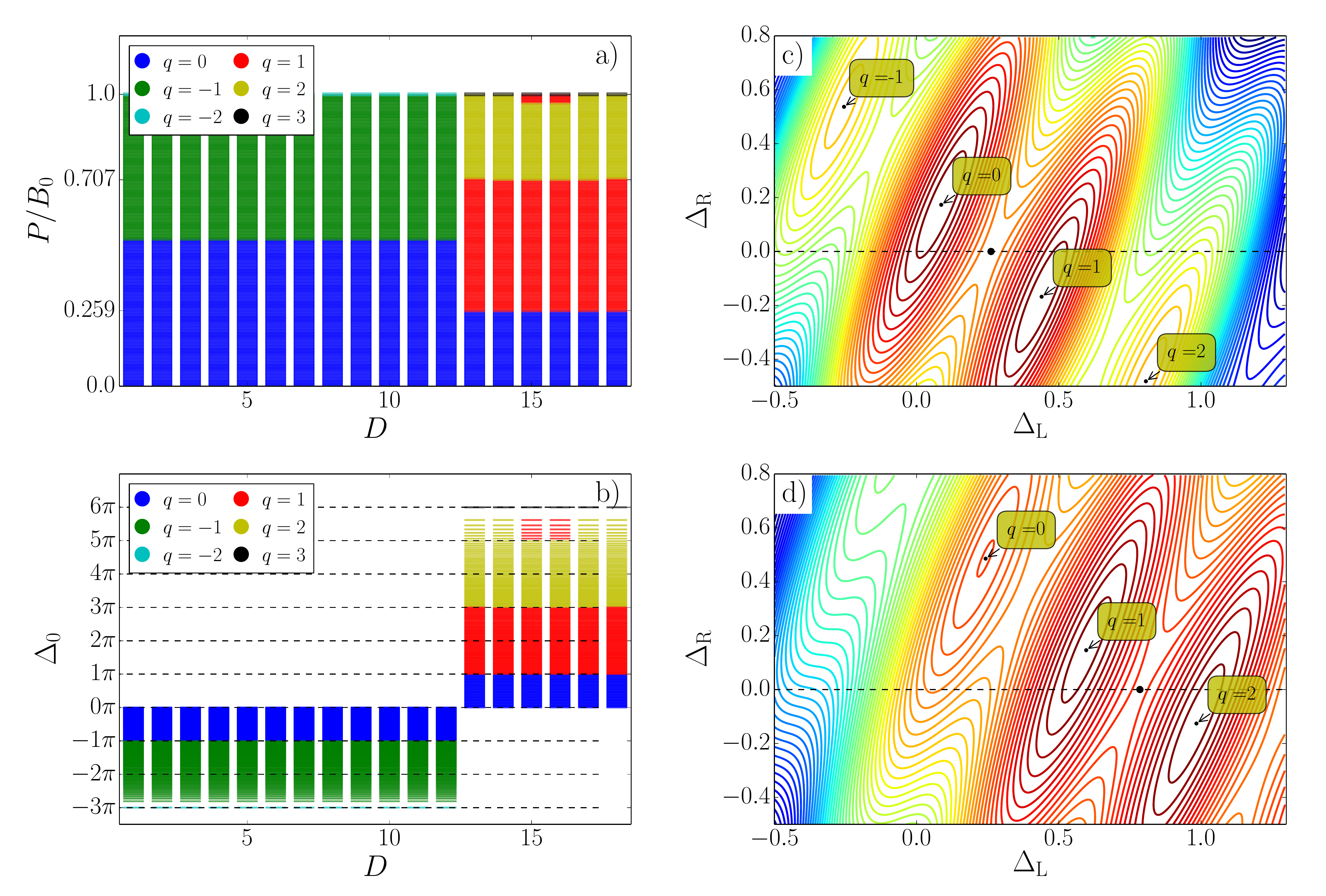}
  \caption{\footnotesize Vortex generation and basins of stability for the single-loop model shown in the 
  inset of Fig.~\ref{fig:Vortex formation by loss of stability}d under the line tripping and reclosing mechanism.
  Panels a) and b): final winding numbers as a function of the position of the 
  tripped line and a) the 
  rescaled injected power $P/B_0$, b) the corresponding angle differences $\Delta_{0}$
  between the two ends of the tripped line. Panels c) and d): 
  Contour plots of the Lyapunov function for a tripped line at $13\leq D\leq18$ (on the right path), 
  c) $P=B_0\sin(\pi/12)$ and d) $P=B_0\sin(\pi/4)$. Local minima with
  different values of $q$ are indicated. 
  When the line is tripped, $\Delta_{\rm R}=0$ and $\Delta_{0}=12 \Delta_{\rm L}$,
  and black dots show the operating states right before reclosing of 
  the tripped line. For the chosen values
  of $P/B_0$ they
  are located precisely on saddle points at the boundary between the basins of attraction 
  of $q=0$ and $q=1$ [panel c)], and $q=1$ and $q=2$ [panel d)].}
  \label{fig:Vortex creation by line reclosing}
 \end{center}
\end{figure}

Fig.~\ref{fig:Vortex creation by line reclosing}c and d show contour plots of 
${\cal V}(\Delta_{\rm L},\Delta_{\rm R})$. Local minima are indicated, together with the
corresponding integer winding numbers. To each minimum corresponds a basin of attraction
containing the set of initial states that converge towards that minimum
under Eq.~(\ref{Swing}). All points around a minimum belong to that basin,
until one reaches a saddle point or a ridge, beyond which points belong to another basin of attraction.
Cutting the right path projects $\Delta_{\rm R} \rightarrow 0$.
Right before line reclosing, the system is 
at $\Delta_{\rm L}= \arcsin(P/B_0)$ on the dashed lines in 
Fig.~\ref{fig:Vortex creation by line reclosing}c and d which 
 correspond to $P/B_0=0.259$ and $P/B_0=0.707$ respectively. The solution 
towards which
the system converges after line reclosing depends on the basin of attraction to which the initial state
belongs. 
For $P$ such that 
$N_{\rm L} \Delta_{\rm L} = \Delta_{0} = (2 p +1) \pi$ with $p \in \mathbb{Z}$, 
the point $(\Delta_L,\Delta_{\rm R})=(\arcsin(P/B_0), 0)$ 
lies right on a saddle-point at the boundary between two basins of attraction, 
as
we now proceed to show.

The gradient of the Lyapunov function 
$\mathcal{V}$ in the $(\Delta_{\rm L},\Delta_{\rm R})$-plane is given by 
\begin{align*}
 \nabla\mathcal{V}&=
 \begin{pmatrix}
  -N_{\rm L}P + N_{\rm L}B_0\sin\Delta_{\rm L}+N_{\rm L}B_0\sin\left(N_{\rm L}\Delta_{\rm L}-(N_{\rm R}-1)\Delta_{\rm R}\right)\\
  (N_{\rm R}-1)B_0\sin\Delta_{\rm R}-(N_{\rm R}-1)B_0\sin\left(N_{\rm L}\Delta_{\rm L}-(N_{\rm R}-1)\Delta_{\rm R}\right)
 \end{pmatrix}\, .
\end{align*}
It is easy to check that $\nabla\mathcal{V}=0$ at $(\Delta_{\rm L},\Delta_{\rm R})=(\arcsin(P/B_0),0)$, which is thus a critical point. 
The nature of this critical point is determined by the two eigenvalues of the Hessian $\mathcal{H}_{\mathcal{V}}$ of $\mathcal{V}$. 
At our critical point, we obtain
\begin{align*}
 \mathcal{H}_{\mathcal{V}}(\arcsin(P/B_0),0)&=B_0
 \begin{pmatrix}
  N_{\rm L}\cos\Delta_{\rm L}-N_{\rm L}^2 
  &  
  N_{\rm L}(N_{\rm R}-1)
  \\
  N_{\rm L}(N_{\rm R}-1) 
  & 
  (N_{\rm R}-1)-(N_{\rm R}-1)^2
 \end{pmatrix}\eqqcolon
 \begin{pmatrix}
  a&b\\
  b&c
 \end{pmatrix}
\, .
\end{align*}
The two eigenvalues of $\mathcal{H}_{\mathcal{V}}$ are then the two roots of 
\begin{align*}
 \chi(\lambda)&=\lambda^2 - (a+c)\lambda + ac-b^2 \implies \lambda^{\pm}=\left(a+c\pm\sqrt{(a+c)^2-4(ac-b^2)}\right)/2\, .
\end{align*}
Then $\lambda^+$ is always positive and $\lambda^-$ is negative if and only if $ac-b^2<0$. Replacing $a$, $b$ and $c$ we have 
\begin{equation}
 ac-b^2=-B_0^2N_{\rm L}(N_{\rm R}-1)(N_{\rm L}+(N_{\rm R}-2)\cos\Delta_{\rm L})\,,
\end{equation}
which is necessarily negative since first, at the moment of the reclosing, 
$\Delta_{\rm L}=\arcsin(P/B_0)$ implying that $\cos\Delta_{\rm L}>0$, and second $N_{\rm R}\geq2$.
We conclude  
that $(\Delta_{\rm L},\Delta_{\rm R})=(\arcsin(P/B_0),0)$ for $N_{\rm L}\arcsin(P/B_0)=(2p+1)\pi$
is a saddle point of the projected Lyapunov function. It can actually be shown that it is a saddle point
of the full Lyapunov function.
One concludes that vortex generation by this mechanism occurs for $ \Delta_0 > \pi$,
and that the final 
winding number increases by one each time $\Delta_0$ crosses an odd
integer multiple of $\pi$.
The same line of argument with $N_{\rm L} \Delta_{\rm L} \leftrightarrow N_{\rm R}\Delta_{\rm R}$
applies when the line to be cut is on the left path.

The above argument is based on the projected Lyapunov function. 
It neglects the fact that, after line reclosing, 
the transient dynamics leaves the $(\Delta_{\rm L},\Delta_{\rm R})$-plane until
a new stationary state is reached. We therefore check its validity numerically.
Fig.~\ref{fig:Vortex creation by line reclosing}
show what final winding number is obtained
upon line reclosing depending on the location $D=1,\dots,n$  
of the open line (counted counterclockwise, starting from 
the main producer) and the rescaled 
power $P/B_0$ (Fig.~\ref{fig:Vortex creation by line reclosing}a) and $\Delta_{0} $
(Fig.~\ref{fig:Vortex creation by line reclosing}b). Fig.~\ref{fig:Vortex creation by line reclosing}b  
confirms that the final 
winding number changes by one each time $\Delta_0$ crosses an odd
integer multiple of $\pi$, except when the injected power gets close to its maximal allowed value, 
$P \rightarrow B_0$. We attribute this change of behavior to a more complicated transient in this case.
Fig.~\ref{fig:Vortex creation by line reclosing}a and b further show that 
$\Delta_0 \approx 2 \pi$ around $P/B_0 \approx 0.5$. Taken modulo $2 \pi$,
this means that the angle difference at the ends of the tripped line is small so that line reclosing is
technically feasible. 

Fig.~\ref{fig:Vortex creation by line reclosing}c) and d) shed a new light on the work of 
Araposthatis et al.~\cite{Ara81}, who discussed the existence of different power flow 
solutions in separated stable domains in voltage angle space in simple models. Our 
method allows to visualize different such domains and in particular to infer the precise
location of saddle points separating them. 
Recent works have advocated a new line of research in dynamical systems including 
coupled oscillators models~\cite{Wil06,Men13} and AC power networks~\cite{Men14},
investigating the size of basins of attraction. These works were
restricted to numerical statistical studies. Our projective approach allows to 
visualize basins of attractions
and the separatrices in between. Quite remarkably, it allows us to understand quantitatively how 
vortices emerge and when winding numbers change. In both line tripping
and line reclosing mechanisms, vortices are created by a topological 
change in the network, which twists the voltage angles around a loop. We therefore 
collectively refer to these two mechanisms as {\it topological phase twist}.

\section{Vortex Flows and Ohmic Dissipation}\label{section6}

We next investigate the persistence of circulating loop flows in the presence of
ohmic dissipation. Voltage angle differences between connected nodes 
in operational states of 
AC power grids seldomly reach more than few tens of degrees, 
beyond which the power line's thermal limit is 
exceeded. Therefore, the dynamical phase slip
mechanism (i) is of little relevance for power grid operation, because it occurs when  
one angle difference
$|\theta_i-\theta_{j}| \gtrsim \pi/2$~\cite{Del16}. We therefore 
focus on the topological phase twist mechanisms (ii) and (iii).

The green and red lines in Fig.~\ref{fig:DoubleLoop} illustrate how ohmic dissipation affects
vortex flow creation by tripping of a line. Because ohmic dissipation requires
an excess of production to compensate for losses, power production is 
$P+\Delta P$, larger
than the power demand $P$. 
One sees that 
the presence of a finite conductance, $\tilde{G}_{lm} \equiv G_0 
\ne 0$ in Eq.~(\ref{eq:pflow_jj_actived}),
reduces the range in $P/B_0$ at which transitions between $q=0$ and $q=1$ occur upon line
tripping, but that the overall behavior remains the same as long as $G_0/B_0$ is 
not too large. Fig.~\ref{fig:DoubleLoop}c shows furthermore how much
more power is consumed in the presence of vortex flows, with a huge stepwise increase in 
ohmic losses by 
almost 50\% of the losses $\Delta P$ in the presence of a still 
moderate conductance $G_0/B_0=0.03$. Fig.~\ref{fig:DoubleLoop} finally shows  
topological protection by the winding number, where once
the vortex has been created, returning the operating conditions to 
smaller $P$ (as indicated by arrows) 
does not bring the system back to the vortex-free state. Instead, the operational
state remains at $q=1$, with losses well above those for $q=0$. 

\begin{figure}
 \begin{center}
  \includegraphics[width=425px]{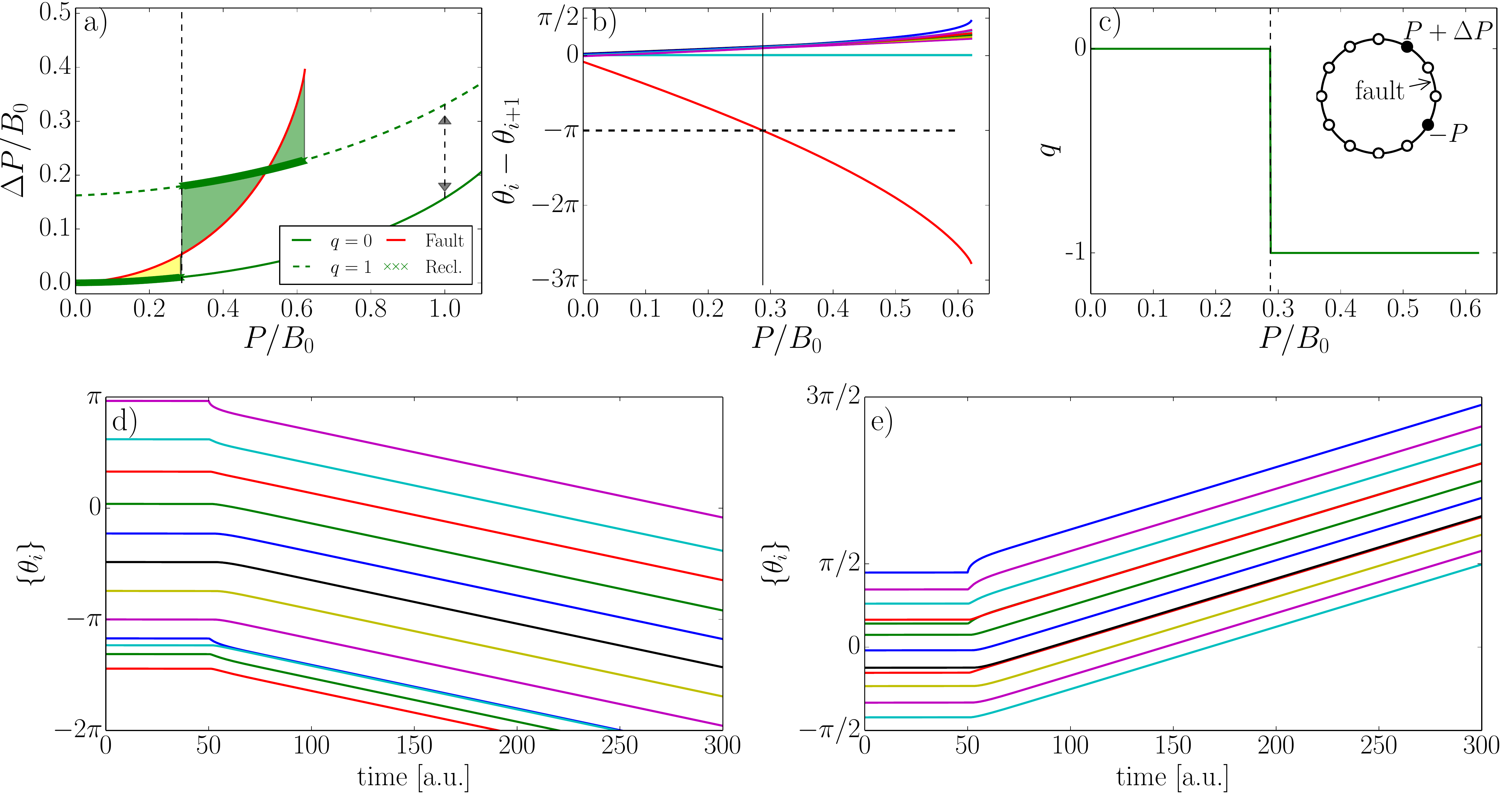}
  \caption{\footnotesize Vortex generation under the line tripping and reclosing mechanism in the presence
  of ohmic dissipation with $G_0/B_0=0.05$ for the model shown in the inset of panel c).
  Panel a): Ohmic losses $\Delta P$ on the closed system with $q=0$ (solid green line)
  and $q=1$ (dashed green line) and on the open loop with one tripped line 
  (red solid line). We start from the $q=0$ solution and cut the line indicated in the inset of panel c).
  Upon closing the line again, the system goes from the red line solution to the $q=0$
  solution in the yellow area, but jumps to $q=1$ in the green area. 
  The red line stops at $P/B_0 \approx 0.62$ above which there is no 
  single-path stable solution.     
  Panel b): angle differences before line reclosing. The jump to $q=1$
  occurs when the angle difference between the two ends of the tripped line 
  exceeds $\pi$, in complete agreement with the basin of attraction argument, 
  even with dissipation. 
  Panel c): final winding number after the tripped line has been reclosed.
  Panels d) and e) show topological protection and robust synchrony with modified frequency as $\Delta P$ is changed as indicated
  by the vertical arrow in panel a) at $P/B_0=1$. Panel d): $\Delta P$ is decreased 
  to $\Delta P_{q=0}$ starting from the $q=1$ state, and e) $\Delta P$ is 
  increased to $\Delta P_{q=1}$ from the $q=0$ state, in both cases at $t=50$. Despite the presence of dissipation and the change in $\Delta P$, the 
  winding number is topologically protected.}
  \label{fig:trip_reclose_ohm}
 \end{center}
\end{figure}

Fig.~\ref{fig:trip_reclose_ohm} makes it clear that vortex generation under the line tripping and reclosing mechanism (iii) proceeds 
 in the same way in the presence of ohmic dissipation. We start from a 
$q=0$ stationary state, cut
a line and let the system relax to a stationary state, after which
we close the line again. What final state is obtained depends on $P/B_0$. When $P$ is small, 
the system relaxes back to the initial state with $q=0$ (yellow area in Fig.~\ref{fig:trip_reclose_ohm}a), however 
at larger $P$, the transient dynamically moves the system towards the $q=1$ stationary state
(green area in Fig.~\ref{fig:trip_reclose_ohm}a). Fig.~\ref{fig:trip_reclose_ohm}b shows that 
the transition to $q=1$ occurs precisely 
when the voltage angle difference between the nodes surrounding the 
faulted line reaches $\pi$, even in the presence of ohmic dissipation, in complete agreement
with the basin of attraction theory discussed above. We finally note that
line reclosing is technically feasible 
around $P/B_0 \approx 0.5$ where the angle difference between the two ends of the tripped line
is small (modulo $2 \pi$). This would lead to a $q=1$ vortex flow state.

Dissipation renders operational conditions different for different stationary
states. In particular, states with vortex flows generically have larger ohmic losses,
because they have larger angle differences in Eq.~(\ref{eq:balance_diss}). They
therefore
require an additional power injection $\Delta P_q$ depending on the vorticity $q$.
One would  think that changing the operational conditions by e.g.
reducing $\Delta P_{q \ne 0} \rightarrow \Delta P_{q = 0}$  makes the $q$-vortex  flow
disappear. Figs.~\ref{fig:trip_reclose_ohm}d and e show however that
topological protection by the winding number 
remains active, despite the presence of dissipation.
In panel d) we decrease the additional power injection $\Delta P$ to the amount of losses incurred
in the $q=0$ stationary state when the system is in the $q=1$ state, while 
in panel e) we increase $\Delta P$ to its value in the $q=1$ state, starting from the $q=0$ state, 
at $P/B_0=1$. In both cases, the winding number remains the same. Synchronization is 
furthermore not  destroyed, however angles rotate at a modified frequency, 
$ \theta_i(t) = \theta_i^{(0)} \pm \delta \Omega t$ in the rotating frame, 
with $\delta \Omega \simeq N^{-1} 
(\Delta P_{q=1}-\Delta P_{q=0})$.
Adapting 
$\Delta P$ to what is required by another $q$-state  changes the  
synchronous frequency but leaves $q$ unchanged. 

\begin{figure}[t]
 \begin{center}
  \includegraphics[width=425px]{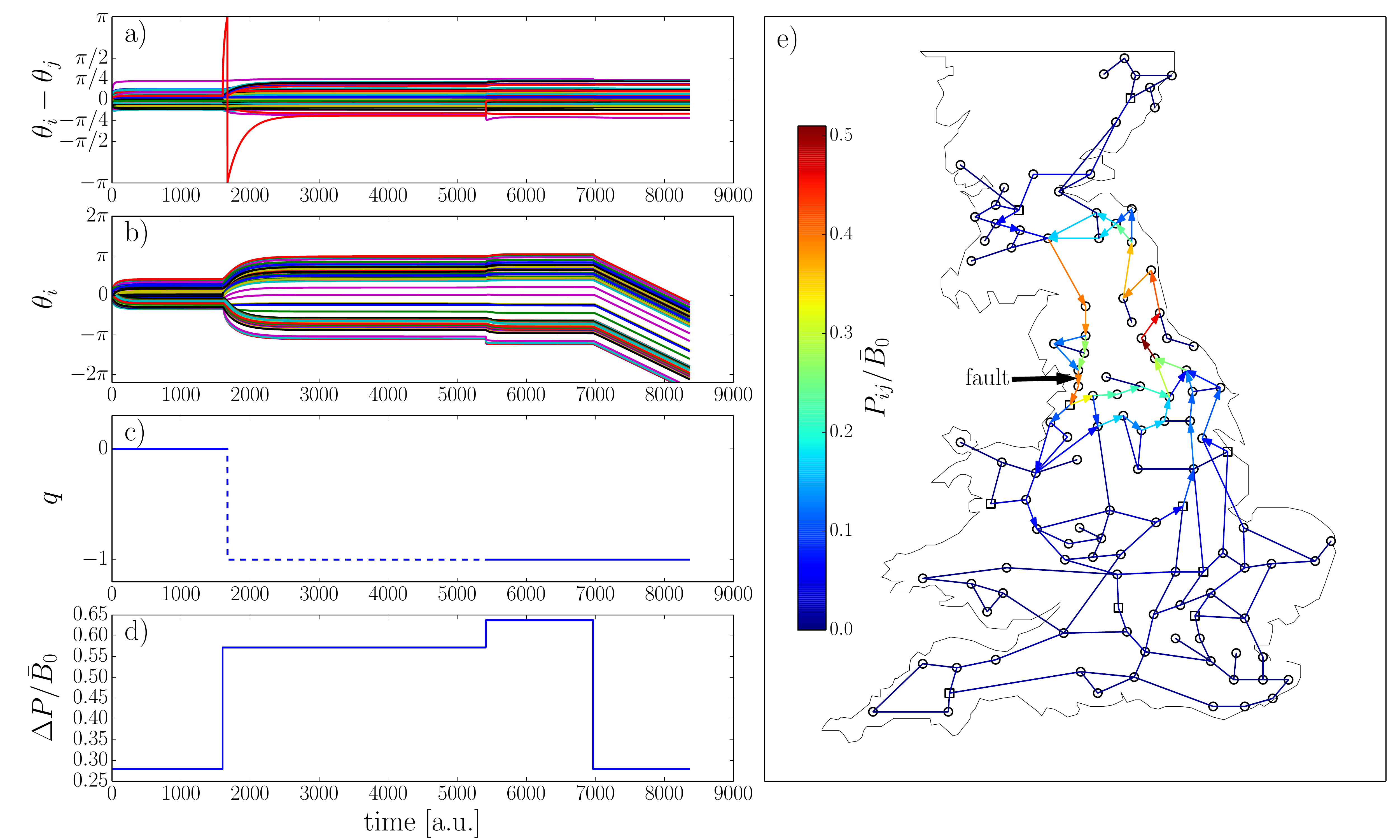}
  \caption{\footnotesize Vortex creation in 
  a complex network with the topology of the UK transmission grid under the line tripping and
  reclosing mechanism. The grid is sketched in panel e). Lines have capacities inversely proportional
  to their length and are normalized to $\langle B_{lm} \rangle = \overline{B}_0$. There are
  10 power generators (indicated by squares ) with 
  $P_{\rm G}= 0.5 \overline{B}_0 + \Delta P/10$
  and 110 consumers (circles) with $P_{\rm C}=-\overline{B}_0/22$.
  Ohmic dissipation is due to a finite conductance with $G_{lm}/B_{lm}=0.1$,
  typical of very high voltage AC power lines.
  Panels a)--d): angle differences, angles, winding number $q$ and 
  ohmic losses as a function of time. We first let the state 
  dynamically converge to a stationary synchronous state without vortex flow. 
  One line then trips at
  $t=1600$ and the system converges to a new synchronous state with increased ohmic losses. 
  The line is reclosed at $t=5410$
  and a vortex flow has been created in the resulting synchronous state, with $q=-1$,
  increasing ohmic losses further. 
  The additional power injected into the grid to compensate for the ohmic
  losses is then brought back to its initial value at $t=6970$.
  The vortex flow persists, angle differences essentially remain unchanged while
  angles start to rotate in unison, $\theta_i = \delta \Omega t$, $\forall i$, 
  indicating a change in 
  synchronous frequency by $\delta \Omega$. The dashed line in panel c) indicate 
  that $q$ is not defined for $t \in [120,330]$, when the loop is open.
  Panel e): color-coded difference in flows between the initial,
  $q=0$ stationary state and the final, $q=-1$ state, in units of $B_{lm}$. 
  Arrows indicate the direction of the flow
  difference only when the latter exceeds 0.05 $\overline{B}_0$.}
  \label{fig:UK}
 \end{center}
\end{figure}

\section{Vortex Flows in Complex Grids}\label{section7}

We finally 
export the  knowledge obtained from investigating simple systems 
to a network model with the topology of the 
UK high-voltage AC power grid~\cite{Col16,Wit12}.
Fig.~\ref{fig:UK} illustrates vortex flow 
creation, enhanced ohmic losses and topological protection of 
stationary states with vortex flows. The system is initially 
stabilized in a stationary state without vorticity on any of its loops. It is later perturbed by a
line tripping, at the position indicated in Fig.~\ref{fig:UK}e. The state is then left
to stabilize towards a new state,
after which it is again perturbed by the reclosing of the line. 
Finally, the power injected is modified to try and move the system back to the stationary state with 
$q=0$ -- without success. Fig.~\ref{fig:UK}a and b show the behavior of the voltage angles
and angle differences between connected nodes. It is seen that line tripping at $t=1600$
essentially makes a single voltage angle difference significantly change, 
similarly to the single-loop model considered 
in Fig.~\ref{fig:trip_reclose_ohm}.
The same angle difference is the only one to move sensibly upon line reclosing at $t=5410$. 
Fig.~\ref{fig:UK}c  shows that the line fault 
creates a vortex flow. The latter is affected by adapting the power at $t=6970$ to the losses
of the initial state with $q=0$ only insofar as all its angles start to rotate at a modified frequency
$\delta \Omega \simeq N^{-1} (\Delta P_{q=0}-\Delta P_{q=-1})$. However this does not affect its
winding number -- topological protection is at work also in this case of a complex meshed grid
with significant ohmic dissipation, $G_{lm}/B_{lm} = 0.1$. 
Fig.~\ref{fig:UK}d shows that the vortex flow doubles
ohmic losses, despite the fact that it affects only a small 
fraction of the grid, as is seen in Fig.~\ref{fig:UK}e which shows differences in 
flows between the $q=-1$ and $q=0$ states. The reduction in losses in Fig.~\ref{fig:UK}d 
after $t=6970$ leads to the synchronization of the grid at a frequency 
different from the rated frequency of 50 Hz. Such a change in frequency would be intolerable in
  a real power grid, and would quickly lead to either controlled or uncontrolled line trippings,
  cascades of failures, possibly leading to blackouts~\cite{Leh10,Vai12,Pah14}. 

\section{Conclusions}\label{section8}

Out of the three mechanisms for creating vortex flows we discussed, the topological 
phase twist
mechanisms (ii) and (iii) are relevant to AC power grid operation as they can occur
at relatively small voltage angle differences between connected nodes. 
The reliability of AC power grids is constantly evaluated via $N-1$ feasibility
and transient stability analysis, where the existence of a stationary solution as well as the convergence
towards that solution is checked after any one of the $N$ major components (lines, transformers etc.) 
is removed from the network. We believe that this analysis should be complemented by 
checks of the presence of vortex flows, since the tripping or the reclosing of a line
has the potential to generate them, resulting in reduced stability and higher, persistent ohmic losses, that are very difficult to get rid of. 

The operating conditions of power grids are expected to change drastically as the energy transition
steadily substitutes smaller, delocalized production for large power plants. As but one consequence,
power generators have less and less mechanical inertia, thus less primary power reserve. 
To compensate for these changes, power electronics devices, phase angle regulating 
transformers and other devices whose task it is to effectively modify admittances and voltage
phase angles are often
incorporated into the power grid. The two 
topological phase twist mechanisms discussed above are in a way extreme, in that 
they rely on line tripping or reclosing. The conditions under which less stringent actions such as
reducing line admittances would generate vortex flows should be investigated. Work along those 
lines is in progress and preliminary results seem to indicate that changes in power grids brought
about by the energy transition have the potential to generate vortex flows more frequently. 

We also discussed vortex flow creation via dynamical phase slip,
despite its lack of relevance for AC power grids. High voltage power transmission 
is however closely connected to problems of coupled oscillators via the 
celebrated Kuramoto model~\cite{Kur84,Str00,Ace05,Are08}. Connections between the 
latter model and Josephson junction arrays were noted in Ref.~\cite{Wie98}.
Stationary states in systems of coupled
oscillators with different winding numbers were discussed in Ref.~\cite{Rog04,Och10}
and in a biological context in Ref.~\cite{Hei15}. Our theory can be exported to those situations
to explain how such states are created in the first place and how they disappear.

We finally comment on further analogies with vortex physics in superconductors. First, the point has
already been made above that there is no counterpart to external magnetic fields in electrical
grids that could generate vortices, and at present it is not known to us whether specific sequences
of injection/consumption changes could lead to the creation of vortex flows. But if such a sequence
exists, Fig.~\ref{fig:Vortex creation by line reclosing} shows that it should move the system over an
"energy" barrier, i.e. over a ridge or a saddle point of the Lyapunov function. This, in a sense,
is to be related to vortex formation in the Ginzburg-Landau model for type II superconductors
where an energy barrier is passed at a critical magnetic 
field beyond which the vortex state is energetically
favorable. Second, vortex creation in superfluids occurs either via creation of a pair of 
vortex-antivortex or via vortex nucleation at the boundary of the system. Electrical power grids 
being rather small, we have found that vortex nucleation occurs at the network boundary, but 
would need to perform numerically intensive investigations on much bigger networks 
(the North American or the Pan-European networks for instance) to see if/when vortex-antivortex pairs
are created. 
While we cannot rule out single vortex creation in the bulk from topological line modifications such as line tripping or reclosing, we suspect this occurs only very rarely, if at all, as such a process would require winding numbers to change simultaneously for a macroscopic number of loops encircling the vortex.
Finally, it is known that vortices in superconductors
move in the presence of a transport current, which leads to dissipation. So far we have seen vortex
flows move only in numerical investigations on regular lattices. That we never saw it in 
complex networks may be due to either the finiteness of the system size considered, or to 
the absence of large loops neighboring those where vortex flows sit or to vortex
pinning in these complex, effectively disordered networks. With our current knowledge,  
whether vortex flows can move around in 
large electric power grids is an open question.

\section*{Acknowledgments}
This work has been supported by the Swiss National Science Foundation under an AP Energy Grant.

\newpage
\renewcommand{\thefigure}{S\arabic{figure}}
\renewcommand{\theequation}{S\arabic{equation}}
\renewcommand{\thesection}{S\arabic{section}}
\setcounter{equation}{0}
\setcounter{section}{0}
\setcounter{figure}{0}
\section*{Topologically Protected Loop flows in High Voltage AC Power Grids: Supplemental Material}

\section{The power flow problem}

Power grids are AC electrical networks. Mathematically speaking, they are 
modeled as graphs with  $n$ nodes, each of them representing a bus 
and the graph's edges representing electrical lines.

The standard operating state of electric power grids 
is characterized by synchrony, where voltage angles on all buses 
rotate at the same frequency.
That state is reached and maintained thanks to the coupling between buses
induced by power lines and a balance regulation between power production and consumption, 
which forbid voltage angles from deviating from a predetermined 
frequency by more than a fraction of a percent. Production and consumption are
constantly fluctuating and accordingly the synchronous state of the system constantly changes.
Most of the time, however, these changes are so slow that the
whole system is effectively in a steady-state determined by one of the solutions to the 
{\it power flow equations},
\begin{subequations}\label{eq:powerflow}
\begin{eqnarray}\label{eq:powerflow_active}
P_l &=& \sum_m  B_{lm} \, |V_l| |V_m| \,  \sin(\theta_l - \theta_m) + G_{lm} \left[ |V_l|^2 - |V_l| |V_m|\cos(\theta_l - \theta_m) \right] \, , \\
Q_l &=& \sum_m  B_{lm} \,   \left[ |V_l|^2 - |V_l| |V_m|\cos(\theta_l - \theta_m) \right] 
- G_{lm} \, |V_l| |V_m| \,  \sin(\theta_l - \theta_m) \, ,
\end{eqnarray}
\end{subequations}
which we take as our starting point.
These are transcendental equations for balanced 3-phase systems~\cite{Ber00}, which
relate the phase and amplitude of the 
complex AC voltage
$V_m = |V_m| \exp[i \theta_m]$ at the bus connecting the $m^{\rm th}$ 
component (a producer or a consumer) 
to the grid, to the active, $P_l$, and reactive, $Q_l$, powers injected ($P_l, Q_l > 0$)
or consumed ($P_l, Q_l < 0$) at 
the $l^{\rm th}$ bus via power lines of complex admittance
$Y_{lm} = G_{lm} + i B_{lm}$. In its full version, the problem is defined by splitting the $n$ buses into
producer and consumer buses
with pre-determined $\{P_l,|V_l |\}$ and $\{P_l,Q_l\}$ respectively. With these conditions fixed,
Eqs.~(\ref{eq:powerflow}) then determine angles at all buses, voltages at consumer buses and
reactive powers at producer buses. 

High voltage power lines 
typically have a conductance to susceptance ratio
$G_{lm}/B_{lm} < 0.1$, becoming smaller at higher voltage. 
A first level of approximation is thus to neglect the conductance
and with it all ohmic dissipation and to consider

\begin{subequations}\label{eq:pflow_jj}
\begin{eqnarray}\label{eq:pflow_jj_active}
P_l &=& \sum_m B_{lm}  |V_l| |V_m| \, \sin(\theta_l - \theta_m)  \, , \\
Q_l &=& \sum_m  B_{lm} \,   \left[ |V_l|^2 - |V_l| |V_m|\cos(\theta_l - \theta_m) \right] \, .
\end{eqnarray}
\end{subequations}
Eqs.~(\ref{eq:pflow_jj}) give the lossless line approximation. 
At that level, one may consider voltage
variations that are necessary to accomodate predetermined amounts of reactive power $Q_l$ at 
consumer nodes. However, a standardly used level of approximation \cite{Fil08,Cho09,Are08} 
is to neglect voltage variations and consider $|V_l| = V_0$, $\forall l$, to introduce an 
effective line susceptance $\tilde{B}_{lm} = B_{lm}  V_0^2$, and to focus on the power flow equation 
for the active power only, Eq.~(\ref{eq:pflow_jj_active}), since, within that approximation, 
it decouples from the reactive power. We will follow that approach.
Within the lossless line approximation, the analogy between AC power grids and
current biased Josephson junction arrays is evident : power flows between buses $l$ and $m$
in AC power grids have the same sinusoidal 
dependence on voltage angle differences, $\tilde{B}_{lm} \sin(\theta_l - \theta_m)$, that Josephson 
currents between two tunnel-coupled superconductors have on the phases of the 
order parameters~\cite{Jos62}.

The fact that the lines are lossless manifests itself in the power balance between production and
consumption, $\sum_l P_l = 0$. When the conductance $G_{lm}$ is not negligible, the power dissipated
must be compensated by an increased power injection. When
dissipation is taken into account, one has
\begin{eqnarray}\label{eq:balance_diss}
\sum_l P_l &=& \sum_{l,m} \tilde{G}_{lm} \, \left[1 - \cos(\theta_l - \theta_m) \right ] \, ,
\end{eqnarray}
where we defined $\tilde{G}_{lm} \equiv G_{lm} V_0^2$. 
Starting from the lossless perspective and considering dissipative effects as a perturbation, 
it is evident from Eq.~\eqref{eq:balance_diss} that different solutions to Eq.~(\ref{eq:pflow_jj_active}) 
dissipate different amounts of power.
Thus, for the full problem (including dissipation), these different solutions generally require different power injections
and the power flow problem is no longer defined a priori by a set of power injections $\{ P_i \}$ which sum up to zero. 
Power production must be greater than the consumption to compensate the ohmic losses.
Thus power injection must be adapted self-consistently with the angles.
In our simulations we will achieve that by increasing/decreasing power injection depending on the frequency of the 
synchronous solution as obtained from the swing equations, Eq.~\eqref{eq:Swing simplified}.
This procedure is iterated until the dynamical system converges toward a synchronous stationary solution
having the reference frequency.
Clearly ohmic losses can be compensated by 
different injection profiles and this procedure is a priori not uniquely defined. A 
standard procedure in electrical engineering is
to predetermine all but one power injections and consumptions and 
introduce a {\it slack bus}, that injects whatever additional power is needed to balance production with
consumption and satisfy Eq.~(\ref{eq:balance_diss}). This is the approach we follow in our simulations on simple network 
topologies presented hereafter and in the main text. 
For the simulation of the complex network having the topology of the UK transmission grid
(see main text) we assume instead that 
all producers equally compensate for the ohmic losses.

\section{Number of different power flow solutions}

Refs.~\cite{Dor13,Del16} showed that different solutions to the dissipationless
power flow Eq.~(\ref{eq:pflow_jj}) exist
on meshed networks, and that these different solutions
are related to one another by circulating loop flows, i.e. power flows going around closed
loops formed in the network. We sketch here the proof of that theorem. It is based on the 
{\it incidence matrix} $A$ of the network, whose row indices correspond to nodes (buses)
and column indices to the network edges (the links between nodes), and which is defined as
\begin{equation}\label{eq:matrix_inc} 
A_{l i} = \left\{
 \begin{array}{ll}
  1\, ,&\text{if node }l\text{ is the source of edge }i\, ,\\
  -1\, &\text{if node }l\text{ is the target of edge }i \, ,\\
  0\, ,&\text{otherwise}\, .
 \end{array}
 \right.
\end{equation}
The dissipationless power flow problem can be rewritten in terms of the incidence matrix as
\begin{align}\label{mlfs_balance2}
 P_l&=\sum_{i}A_{li} \,I_i \, ,
\end{align}
where $I_i$ is a component of the vector of flows on the network's edges.
It is easy to see that two solutions of Eq.~(\ref{mlfs_balance2}) differ by a flow vector 
with components $\delta I_i$ satisfying
\begin{align}\label{mlfs_balance_diff}
 0 &=\sum_{i}A_{li} \, \delta I_i \, ,
\end{align}
and which therefore belongs to the kernel of the incidence matrix. The proof is completed by invoking a 
known result of algebraic graph theory that any element in $\ker(A)$ is a linear combination 
of unitary flows along loops formed in the network~\cite{Big93}.
Reference~\cite{Del16} also discusses how these different solutions can be labeled by integer topological winding numbers.

We qualitatively discuss the fate of circulating loop flows in cases with dissipation, i.e. nonegligible
conductance. From Eq.~(\ref{eq:matrix_inc}), for each element of the incidence matrix 
$A_{li} \ne 0$ there is exactly one $A_{mi} = -A_{li}$, where $l$ and $m$ label the two ends of
edge $i$. Thus contributions from each nondissipative, 
susceptive flow $I_i$ appear for both ends of edge $i$ in Eq.~(\ref{mlfs_balance2}), 
contributing to $P_l$ and $P_m$ with opposite sign. 
This parity antisymmetry is characteristic of nondissipative flows.
On the contrary, the dissipative part of the flow is parity symmetric - a current of fixed magnitude 
on a given edge gives the same dissipation, regardless of the
direction in which it traverses the line. Adding dissipation therefore requires to add a {\it resistance}
matrix $R$ with $R_{li}= |A_{li}|$.
Eq.~(\ref{mlfs_balance2}) becomes, in the presence of dissipation
\begin{align}\label{mlfs_balance2_dissipation}
 P_l&=\sum_{i}A_{li}\,I_i + R_{li}\,J_i\, ,
\end{align}
with a dissipated power flow $J_i \ge 0$. The difference between two power flow solutions
then reads 
\begin{align}\label{difference_ohmic}
 \delta P_l&=\sum_{i}A_{li} \,\delta I_i + R_{li} \,\delta J_i\, .
\end{align}
Summing over all nodes one gets 
\begin{align}\label{sum_difference_ohmic}
\sum_l \delta P_l&=\sum_{l,i} R_{li} \,\delta J_i\, ,
\end{align}
which, similarly as Eq.~(\ref{eq:balance_diss}), says that 
the net total injected power is solely due to  the dissipative part of the power flow. 
Assuming that the equality in Eq.~(\ref{sum_difference_ohmic}) holds term by term, 
one obtains again Eq.~(\ref{mlfs_balance_diff}) so that 
the two solutions differ by loop flows only. How this can be achieved in practice is easily seen
starting from the lossless line approximation, Eq.~(\ref{eq:pflow_jj}), which, as was shown in 
Refs.~\cite{Dor13,Del16}, carries solutions $\{\theta_i^{(0)}\}$ differing 
by loop flows only. Once dissipation is added, these solutions remain valid if one chooses to 
change the injected and consumed powers as $P_l \rightarrow P_l + \delta P_l$ with 
\begin{eqnarray}\label{eq:balance_diss_diff}
\delta P_l &=& \sum_m \bar{G}_{lm} \, \left[1 - \cos(\theta_l^{(0)} - \theta_m^{(0)}) \right ] \, .
\end{eqnarray}
This suggests that, at least for not too strong dissipation, i.e. for weak conductance to susceptance
ratios, different solutions to the power flow problem exist. 
For two such solutions, the transmitted powers still differ by loop flows.
Extending the results of Refs.~\cite{Dor13,Del16} is straightforward with the 
choice of Eq.~(\ref{eq:balance_diss_diff}) for compensating losses, 
because angles are not modified. In the main text, we numerically use another procedure, where
the additional power necessary to counterbalance ohmic losses is injected from a single bus. 
This is a standard approach in electrical engineering where a {\it slack bus} ensures
that the total power balance is satisfied~\cite{Ber00}. 

Because the power flow problem in the presence of dissipation becomes higher-dimensional,
with additional degrees of freedom related to the choice of producers in charge of compensating the ohmic
dissipation, it is much harder to make general statements about the extension of the theorem of Refs.~\cite{Dor13,Del16} to that case. 
Nevertheless, from Eq.~(\ref{eq:balance_diss_diff}) it is clear that since solutions indexed by high topological 
winding numbers (i.e. solutions carrying large loop flows) are characterized by larger phase differences, they will in general lose more power to ohmic losses.

\section{Dynamics and stability}

For each producer  and  consumer bus, energy conservation states that the injected or consumed
power is equal to the transmitted power (with a negative or positive sign depending on whether it
flows toward or away from the bus considered) minus the losses, including transmission line losses 
as well as mechanical damping losses. 
This balance condition leads to 
the {\it swing equations}~\cite{Ber00}
\begin{equation}\label{Swing}
M_l\ddot\theta_l+\dot{\theta_l}=P_l-  \sum_m  \left(B_{lm} \, |V_l| |V_m| \,  \sin(\theta_l - \theta_m) + G_{lm} \left[ |V_l|^2 - |V_l| |V_m|\cos(\theta_l - \theta_m) \right]\right) \, ,
\end{equation}
which we write here in a frame rotating with the frequency $\Omega/2 \pi = 50$ or 60 Hz of the grid.
The terms $M_l\ddot \theta_l$ and $\dot\theta_l$ in Eq.~\eqref{Swing} represent the change in rotational kinetic energy 
and the damping of the rotating machines connected to the grid.

In this work we consider a simplified version of the 
swing equations where the mechanical inertia of generators is neglected, 
i.e. where instead of Eq.~\eqref{Swing} we consider
\begin{equation}\label{eq:Swing simplified}
\dot{\theta_l}=P_l- \sum_m \tilde{B}_{lm}\, \sin(\theta_l - \theta_m) + \tilde{G}_{lm}\left[1-\cos(\theta_l-\theta_m)\right]\,.
\end{equation}
While the inertia term
affects stability time scales~\cite{Ber00}, it does not influence whether a solution is linearly 
stable or not, which is our interest here. 
In particular it can be shown that linear stability is lost for 
a power grid modeled by Eq.~(\ref{eq:Swing simplified}) at the same time it would be lost for the same set
of equations extended with inertia terms with any distribution of inertia $\{M_l\}$~\cite{Col16,Man14}.
We therefore neglect inertia terms from now on.

Solutions to the power flow equations, Eq.~(\ref{eq:powerflow}), are 
stationary solutions of the swing equations, Eq.~(\ref{eq:Swing simplified}).
The latter allows to determine the
linear stability of power flow solutions $\{\theta_l^{(0)}\}$ under small perturbations, 
$\theta_l^{(0)} \rightarrow \theta_l^{(0)}
+ \delta \theta_l$. Within the lossless line approximation, the linearized dynamics reads
\begin{equation}\label{SwingLin}
\delta \dot{\theta_l}=-\sum_{m} \tilde{B}_{lm} \, \cos(\theta_l^{(0)}-\theta_m^{(0)}) 
(\delta \theta_l-\delta \theta_m) \, .
\end{equation}
The linear stability of the solution  is therefore determined 
by the spectrum of the {\it stability matrix} $M(\{\theta_l^{(0)}\})$, 
\begin{equation}\label{stabilityM}
 M_{lm} = \left\{
 \begin{array}{cl}
 \tilde{B}_{lm}  \cos(\theta_l^{(0)}-\theta_m^{(0)})&\quad\text{if }l\neq m\, ,\\
  \displaystyle -\sum_{k\neq l} \tilde{B}_{lk} \cos(\theta_l^{(0)}-\theta_k^{(0)})&\quad\text{if }l=m\, ,
  \end{array}
  \right.
\end{equation}
which depends on the angles at the stationary, phase-locked solution. 
The eigenvalues of $M$ are the so-called Lyapunov exponents~\cite{Pai89}.
Without dissipation, $M$ is real symmetric, therefore all Lyapunov exponents are real,
furthermore one of them always vanishes, $\lambda_1 = 0$, because 
$\sum_j M_{ji}  = \sum_j M_{ij} = 0$. This condition is similar to a gauge invariance 
according to which 
only angle differences matter.
The stationary state is thus linearly stable if $M$ 
is negative semidefinite and unstable otherwise. 
Equivalently, stability is ensured as long
as the 
largest nonvanishing eigenvalue $\lambda_2$ of $M$ remains negative.

In the dissipative case, the definition of the stability matrix generalizes to 
\begin{equation}\label{eq:stabilityM dissipation}
 M_{lm} = \left\{
 \begin{array}{cl}
 \tilde{B}_{lm}  \cos(\theta_l^{(0)}-\theta_m^{(0)})-\bar{G}_{lm}  \sin(\theta_l^{(0)}-\theta_m^{(0)}) &\quad\text{if }l\neq m\, ,\\
  \displaystyle -\sum_{k\neq l} \left(\tilde{B}_{lk} \cos(\theta_l^{(0)}-\theta_k^{(0)})-\bar{G}_{lk} \sin(\theta_l^{(0)}-\theta_k^{(0)})\right) &\quad\text{if }l=m\,.
  \end{array}
  \right.
\end{equation}
Thus, in the presence of dissipation, the stability matrix maintains its zero row sum property (ensuring that $\lambda_1=0$)
but is no longer symmetric which can lead to a complex spectrum. In this case linear stability is ensured as long
as the real part of all nonvanishing eigenvalues of $M$ is negative. To make contact with the 
nondissipative case,
we denote by $\lambda_2$ the eigenvalue having the largest real part.

\section{Dynamical generation of vortices}

The first mechanism for generating vortex flows discussed in the main text is based on the 
loss of stability of power flow solutions with lower winding numbers while solutions with 
higher winding numbers remain stable. This is done by increasing power generated and consumed,
which, depending on the grid's geometry leads to a line congestion and a temporary 
dynamical instability, eventually driving the system to a new, stable stationary state with 
redistributed power flows. The model considered in the main text is a ring of $n=18$ nodes, with 
one main producer at node 1 and one main consumer at node 7. Small random injections
at all other nodes are introduced to 
remove a mirror symmetry  along the axis going through nodes 4 and 13. This 
symmetry results in $\dot{\theta}_i = -\dot{\theta}_{{\cal P}(i)}$ with ${\cal P}(i)$ denoting the 
mirror symmetric node to $i$. Analytically, this forbids transitions to other $q$-values starting
from any stationary $q=0$ states. Numerically it results in very long transients with 
anomalously long angle rotations. To show that our model is generic and that its behavior is
not an artefact of random injections/consumptions we discuss a different model which 
corroborates our conclusions in the main text. 

\begin{figure}[t]
 \begin{center}
  \includegraphics[width=0.6\textwidth]{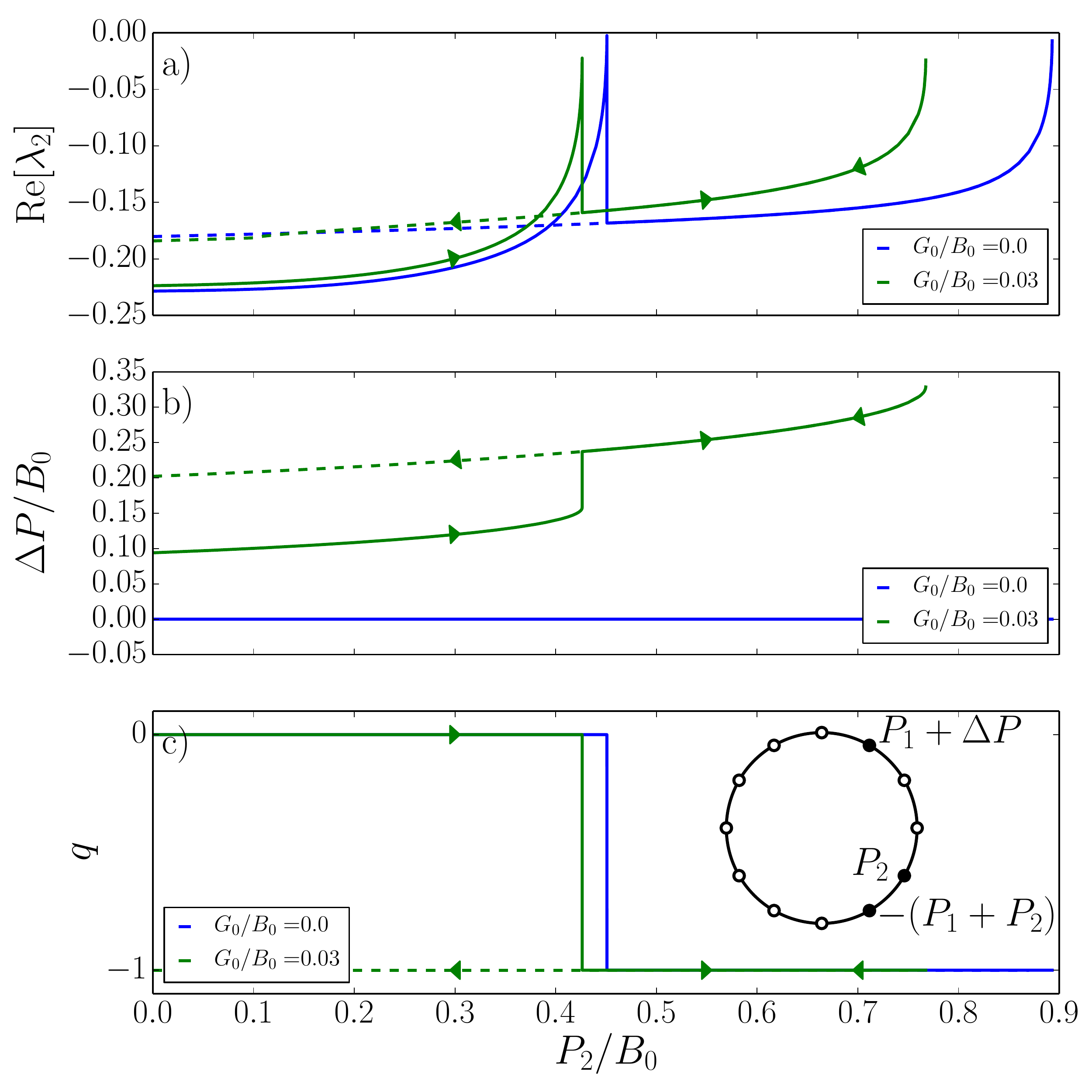}
  \caption{Dynamical generation of vortices by line weakening for a ring with $n=12$ nodes and lines of susceptance $B_0$. 
   Locations of the main power injections and consumptions are indicated by the black nodes in the inset of panel c), while all white nodes have small
   random injections and consumptions summing to zero, to make the model more generic.
   As the consumer demand is increased from $-P_1=-B_0$ to $-P_1-P_2$ units of power, the transmission path connecting the comsumer to its neighboring 
   producer weakens forcing additional power to flow around the other side of the loop.
   As the power demand is increased beyond $P_2\simeq0.45B_0$ for the lossless case, and $P_2\simeq0.42B_0$ for the dissipative case $G_0/B_0=0.03$, 
   the $q=0$ solution becomes unstable [panel a)] driving the system into a state with winding number $q=-1$. This state is highly dissipative and, as indicated
   by the arrows, it is topologically protected.}
  \label{figS1}
 \end{center}
\end{figure}

In Fig.~\ref{figS1} 
we illustrate a similar mechanism based on the weakening of a line on 
a network loop by an additional power injection.
We consider a single loop of $n=12$ nodes and lines of capacity $B_0$.
Initially, a  first producer supplies $P_1=B_0$ units of power to a consumer located four nodes away from it.
The transmitted power splits over the two possible paths joining the producer to the consumer.
Next, the consumption is increased from $-P_1$ to $-P_1-P_2$.
To meet this additional power demand, a second generator, neighboring the consumer, injects a power $P_2>0$ 
(see inset of Fig.~\ref{figS1}).
This additional power injection increases the load of the line connecting the consumer to the second generator.
This weakening of the transmission path eventually drives part of the injection of the first producer
away from its original path, around the other side of the loop. 
Fig.~\ref{figS1} illustrates this mechanism respectively in the lossless and dissipative cases where, for the latter, 
the first generator is in charge of fully compensating the ohmic losses $\Delta P$.
The $q=0$ solution loses stability when the additional power injection reaches $P_2\simeq 0.45B_0$ (resp. $P_2\simeq 0.42B_0$ for $G_0/B_0=0.03$),
which drives the system to the $q=-1$ solution. That solution remains stable until $P_2\simeq 0.9B_0$ 
(resp. $P_2\simeq 0.77B_0$ for $G_0/B_0=0.03$). In the dissipative case, the $q=-1$ solution significantly increases ohmic losses 
doubling them at $P_2=0$.
The arrows indicate the hysteretic behavior where
decreasing $P_2$ from the $q=-1$ solution at $P_2\simeq0.77B_0$ does not bring the system
back to the $q=0$ solution. This topological protection forces the power system to produce
more to compensate for the additional ohmic losses despite the existence of a less dissipative solution. 
When the system is in the $q=-1$ state for $P_2\lesssim 0.42B_0$, reducing $\Delta P$ to the value required 
by the $q=0$ solution drives the system to a $q=-1$ synchronous state with
reduced frequency, but not to the $q=0$ state.

\begin{figure}[t]
 \begin{center}
  \includegraphics[width=425px]{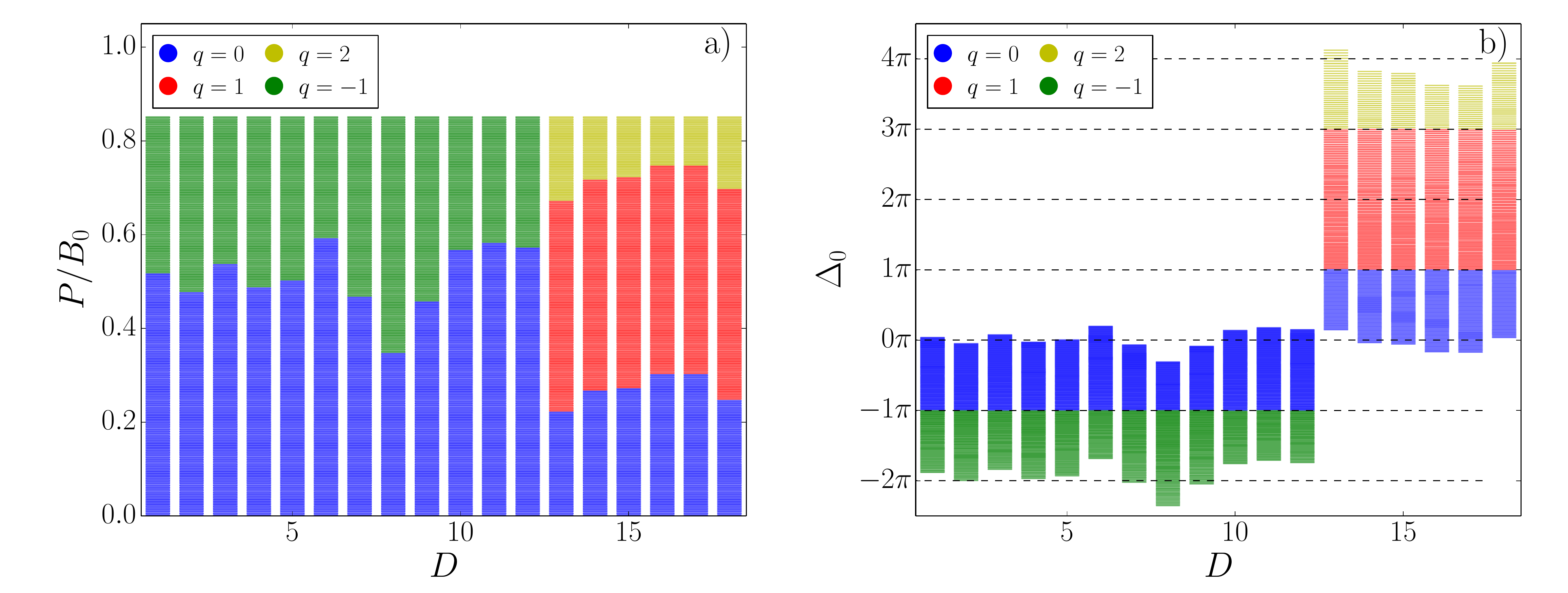}
  \caption{Vortex generation under the line reclosing mechanism, for the single-loop model 
  shown in the inset of Fig.~1d in the main text, with additional random power injection in the intermediate
  nodes.
  The final winding numbers, after reclosing, are color-coded and plotted
  as a function of the position $D$ of the tripped line and a) the 
  rescaled injected power $P/B_0$, b) the angle differences $\Delta_0$
  between the two ends of the tripped line. Vortex generation occurs as soon as
  $\Delta_0 \ge \pi$ and higher winding numbers $q$ are reached each time
  $\Delta_0$ crosses an odd integer multiple of $\pi$. This shows that
  the argument presented in the main text holds even in the more general situation with
  power injections between the
  main producer and the main consumer.}
  \label{figS2}
 \end{center}
\end{figure}

\section{Generation of vortices by line reclosing }

The line reclosing mechanism discussed in the main text considers a big producer connected to a
big consumer on a loop where all other buses have $P_l=0$. This allows to project the $(n-1)$-dimensional configuration
space of angles on a two-dimensional space and in this way to visualize the basins of attraction
of solutions with different winding numbers (Fig.~2c and d of the main text). It was found that the winding numbers change by one
each time the angle difference $\Delta_0$ between the buses at the two ends of the line to
be reclosed crosses an odd integer multiple of $\pi$. 
These findings are generic,
in that they remain valid under the addition of randomly distributed power injections/consumptions on 
intermediate buses between the big producer and consumer. This is illustrated in Fig.~\ref{figS2}.
Compared to Fig.~2 of the main text, the presence of random injections/consumptions strongly modifies 
the borders between reconnected solutions with different $q$ in the $P/B$ vs. $D$ plane (panel a) but not in the $\Delta_0$ vs. $D$ plane (panel b). 
This is so because the small random injections change the value of $P$ necessary for $\Delta_0=(2p+1)\pi$.
One sees that, despite the presence of random power injections, 
vortex flows are generated by line reconnection as soon as $|\Delta_0|>\pi$, but not earlier, 
and that the created vortex
has a winding number increasing/decreasing as $\Delta q = {\rm mod}(\Delta_0+\pi, 2 \pi)$
with good precision. This confirms that the findings presented in the main text are generic and not
restricted to the a priori ideal situation considered there.

\section{Basins of attraction and the projected Lyapunov function}
In this section we present some details on the basins
of attraction of solutions having different winding numbers in the context of the line tripping and reclosing
mechanism for the model depicted in the inset of Fig.~1d of the main text.

In the main text, we show that the Lyapunov function projected 
on the $(\Delta_{\rm L},\Delta_{\rm R})$-plane reads,
\begin{align}
 {\cal V}(\Delta_{\rm L},\Delta_{\rm R})&=-N_{\rm L}P\Delta_{\rm L}-N_{\rm L}B_0\cos\Delta_{\rm L}-(N_{\rm R}-1)B_0\cos\Delta_{\rm R}-B_0\cos(\Delta_0)\nonumber\\
 &=-N_{\rm L}P\Delta_{\rm L}-N_{\rm L}B_0\cos\Delta_{\rm L}-(N_{\rm R}-1)B_0\cos\Delta_{\rm R}-B_0\cos(N_{\rm L}\Delta_{\rm L}-(N_{\rm R}-1)\Delta_{\rm R})\, . \label{si_eq:lyapfunctionplane}
\end{align}
This is Eq.~(9) in the main text.
Also in the main text, we show that the point $(\Delta_{\rm L},\Delta_{\rm R})=(\arcsin(P/B_0),0)$ 
(the system's state just before  
reclosing of the faulty line) is a saddle point of the Lyapunov function \eqref{si_eq:lyapfunctionplane}, 
if $P$ is such that $N_{\rm L}\Delta_{\rm L}=(2p+1)\pi$ for $p\in\mathbb{Z}$. 
Inspection of the Hessian of the Lyapunov function allows us further to 
show that $(\Delta_{\rm L},\Delta_{\rm R})=(\arcsin(P/B_0),0)$ for $N_{\rm L}\arcsin(P/B_0)=(2p+1)\pi$
is a saddle point of the projected Lyapunov function.
We can thus tune parameters precisely in such a way that when the line
is reclosed, one gets a final operating state with the desired vorticity. 

\begin{figure}[t]
 \begin{center}
  \includegraphics[width=425px]{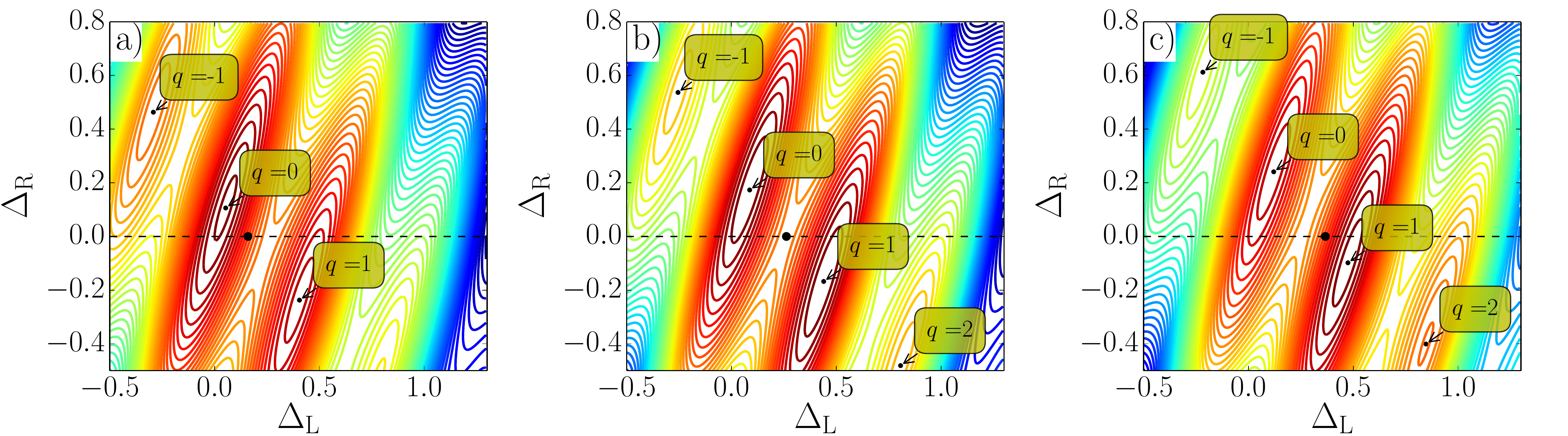}
  \caption{Contour plots of the projected Lyapunov function for the 18 node cycle network described in the main text [see Fig.~1d)]
  and injected powers $P\approx0.159B_0$ panel a), $P=B_0\sin(\pi/12)\approx0.259B_0$ panel b), and $P\approx0.359B_0$ panel c).
  Local minima at different values of $q$ are indicated. 
  The operating state obtained by tripping one line on the shortest path between the generator and the producer 
  ($\Delta_{\rm R}=0$, $\Delta_{\rm L}=\arcsin{(P/B_0)}$, and $\Delta_0=12\Delta_{\rm L}$) 
  is indicated by the black dot on each panel. It lies in the basin of attraction of the $q=0$ state for $P<0.259B_0$ [case a)], 
  in the basin of atttraction of the $q=1$ state for $P>0.259B_0$ [case c)], 
  and at the saddle point separating the two basins for $P=0.259B_0$ [case b)].}
  \label{fig:Basins}
 \end{center}
\end{figure}

Fig.~\ref{fig:Basins} gives a contour plot of the Lyapunov function for the $18$ node cycle network considered in the main text which illustrates how the final vorticity can be tuned. 
Depending on the value of the power injected [panels a) to c)], the state of the system at the moment of the line reclosing 
lies either in the basin of attraction of the $q=0$ solution, or in that of the $q=1$ solution, 
or at a saddle point separating them.

\bibliographystyle{apsrev4-1}
\bibliography{biblio_mlfs}

\begin{thebibliography}{45}%
\makeatletter
\providecommand \@ifxundefined [1]{%
 \@ifx{#1\undefined}
}%
\providecommand \@ifnum [1]{%
 \ifnum #1\expandafter \@firstoftwo
 \else \expandafter \@secondoftwo
 \fi
}%
\providecommand \@ifx [1]{%
 \ifx #1\expandafter \@firstoftwo
 \else \expandafter \@secondoftwo
 \fi
}%
\providecommand \natexlab [1]{#1}%
\providecommand \enquote  [1]{``#1''}%
\providecommand \bibnamefont  [1]{#1}%
\providecommand \bibfnamefont [1]{#1}%
\providecommand \citenamefont [1]{#1}%
\providecommand \href@noop [0]{\@secondoftwo}%
\providecommand \href [0]{\begingroup \@sanitize@url \@href}%
\providecommand \@href[1]{\@@startlink{#1}\@@href}%
\providecommand \@@href[1]{\endgroup#1\@@endlink}%
\providecommand \@sanitize@url [0]{\catcode `\\12\catcode `\$12\catcode
  `\&12\catcode `\#12\catcode `\^12\catcode `\_12\catcode `\%12\relax}%
\providecommand \@@startlink[1]{}%
\providecommand \@@endlink[0]{}%
\providecommand \url  [0]{\begingroup\@sanitize@url \@url }%
\providecommand \@url [1]{\endgroup\@href {#1}{\urlprefix }}%
\providecommand \urlprefix  [0]{URL }%
\providecommand \Eprint [0]{\href }%
\providecommand \doibase [0]{http://dx.doi.org/}%
\providecommand \selectlanguage [0]{\@gobble}%
\providecommand \bibinfo  [0]{\@secondoftwo}%
\providecommand \bibfield  [0]{\@secondoftwo}%
\providecommand \translation [1]{[#1]}%
\providecommand \BibitemOpen [0]{}%
\providecommand \bibitemStop [0]{}%
\providecommand \bibitemNoStop [0]{.\EOS\space}%
\providecommand \EOS [0]{\spacefactor3000\relax}%
\providecommand \BibitemShut  [1]{\csname bibitem#1\endcsname}%
\let\auto@bib@innerbib\@empty
\bibitem [{\citenamefont {Casazza}(1998)}]{Cas98}%
  \BibitemOpen
  \bibfield  {author} {\bibinfo {author} {\bibfnamefont {J.}~\bibnamefont
  {Casazza}},\ }\href@noop {} {\bibfield  {journal} {\bibinfo  {journal}
  {Electrical World}\ }\textbf {\bibinfo {volume} {212}},\ \bibinfo {pages}
  {62} (\bibinfo {year} {1998})}\BibitemShut {NoStop}%
\bibitem [{\citenamefont {Lerner}(2003)}]{Ler03}%
  \BibitemOpen
  \bibfield  {author} {\bibinfo {author} {\bibfnamefont {E.~J.}\ \bibnamefont
  {Lerner}},\ }\href@noop {} {\bibfield  {journal} {\bibinfo  {journal} {The
  Industrial Physicist}\ }\textbf {\bibinfo {volume} {9}},\ \bibinfo {pages}
  {8} (\bibinfo {year} {2003})}\BibitemShut {NoStop}%
\bibitem [{\citenamefont {Whitley}(2008)}]{Whi08}%
  \BibitemOpen
  \bibfield  {author} {\bibinfo {author} {\bibfnamefont {S.~G.}\ \bibnamefont
  {Whitley}},\ }\href@noop {} {\emph {\bibinfo {title} {Lake Erie Loop Flow
  Mitigation}}}\ (\bibinfo  {publisher} {Technical Report; New York Independent
  System Operator},\ \bibinfo {year} {2008})\BibitemShut {NoStop}%
\bibitem [{\citenamefont {Onsager}(1949)}]{Ons49}%
  \BibitemOpen
  \bibfield  {author} {\bibinfo {author} {\bibfnamefont {L.}~\bibnamefont
  {Onsager}},\ }\href@noop {} {\bibfield  {journal} {\bibinfo  {journal} {Nuovo
  Cimento}\ }\textbf {\bibinfo {volume} {6}},\ \bibinfo {pages} {249} (\bibinfo
  {year} {1949})}\BibitemShut {NoStop}%
\bibitem [{\citenamefont {Feynman}(1955)}]{Fey55}%
  \BibitemOpen
  \bibfield  {author} {\bibinfo {author} {\bibfnamefont {R.~P.}\ \bibnamefont
  {Feynman}},\ }\href@noop {} {\bibfield  {journal} {\bibinfo  {journal}
  {Progress in Low Temperature Physics}\ }\textbf {\bibinfo {volume} {1}},\
  \bibinfo {pages} {34} (\bibinfo {year} {1955})}\BibitemShut {NoStop}%
\bibitem [{\citenamefont {Byers}\ and\ \citenamefont {Yang}(1961)}]{Bye61}%
  \BibitemOpen
  \bibfield  {author} {\bibinfo {author} {\bibfnamefont {N.}~\bibnamefont
  {Byers}}\ and\ \bibinfo {author} {\bibfnamefont {C.~N.}\ \bibnamefont
  {Yang}},\ }\href@noop {} {\bibfield  {journal} {\bibinfo  {journal} {Phys.
  Rev. Lett.}\ }\textbf {\bibinfo {volume} {7}},\ \bibinfo {pages} {46}
  (\bibinfo {year} {1961})}\BibitemShut {NoStop}%
\bibitem [{\citenamefont {Bergen}\ and\ \citenamefont {Vittal}(2000)}]{Ber00}%
  \BibitemOpen
  \bibfield  {author} {\bibinfo {author} {\bibfnamefont {A.~R.}\ \bibnamefont
  {Bergen}}\ and\ \bibinfo {author} {\bibfnamefont {V.}~\bibnamefont
  {Vittal}},\ }\href@noop {} {\emph {\bibinfo {title} {Power {Systems}
  {Analysis}}}}\ (\bibinfo  {publisher} {Prentice Hall},\ \bibinfo {year}
  {2000})\BibitemShut {NoStop}%
\bibitem [{\citenamefont {Josephson}(1962)}]{Jos62}%
  \BibitemOpen
  \bibfield  {author} {\bibinfo {author} {\bibfnamefont {B.~D.}\ \bibnamefont
  {Josephson}},\ }\href@noop {} {\bibfield  {journal} {\bibinfo  {journal}
  {Phys. Lett.}\ }\textbf {\bibinfo {volume} {1}},\ \bibinfo {pages} {251}
  (\bibinfo {year} {1962})}\BibitemShut {NoStop}%
\bibitem [{\citenamefont {D\"orfler}\ \emph {et~al.}(2013)\citenamefont
  {D\"orfler}, \citenamefont {Chertkov},\ and\ \citenamefont {Bullo}}]{Dor13}%
  \BibitemOpen
  \bibfield  {author} {\bibinfo {author} {\bibfnamefont {F.}~\bibnamefont
  {D\"orfler}}, \bibinfo {author} {\bibfnamefont {M.}~\bibnamefont {Chertkov}},
  \ and\ \bibinfo {author} {\bibfnamefont {F.}~\bibnamefont {Bullo}},\
  }\href@noop {} {\bibfield  {journal} {\bibinfo  {journal} {Proc. Natl. Acad.
  Sci.}\ }\textbf {\bibinfo {volume} {110}},\ \bibinfo {pages} {2005} (\bibinfo
  {year} {2013})}\BibitemShut {NoStop}%
\bibitem [{\citenamefont {Delabays}\ \emph {et~al.}(2016)\citenamefont
  {Delabays}, \citenamefont {Coletta},\ and\ \citenamefont {Jacquod}}]{Del16}%
  \BibitemOpen
  \bibfield  {author} {\bibinfo {author} {\bibfnamefont {R.}~\bibnamefont
  {Delabays}}, \bibinfo {author} {\bibfnamefont {T.}~\bibnamefont {Coletta}}, \
  and\ \bibinfo {author} {\bibfnamefont {P.}~\bibnamefont {Jacquod}},\
  }\href@noop {} {\bibfield  {journal} {\bibinfo  {journal} {J. Math. Phys.}\
  }\textbf {\bibinfo {volume} {57}},\ \bibinfo {pages} {032701} (\bibinfo
  {year} {2016})}\BibitemShut {NoStop}%
\bibitem [{\citenamefont {Taylor}(2012)}]{Tay12}%
  \BibitemOpen
  \bibfield  {author} {\bibinfo {author} {\bibfnamefont {R.}~\bibnamefont
  {Taylor}},\ }\href@noop {} {\bibfield  {journal} {\bibinfo  {journal} {J.
  Phys. A}\ }\textbf {\bibinfo {volume} {45}},\ \bibinfo {pages} {055102}
  (\bibinfo {year} {2012})}\BibitemShut {NoStop}%
\bibitem [{\citenamefont {Mehta}\ \emph {et~al.}(2015)\citenamefont {Mehta},
  \citenamefont {Daleo}, \citenamefont {D\"orfler},\ and\ \citenamefont
  {Hauenstein}}]{Meh14}%
  \BibitemOpen
  \bibfield  {author} {\bibinfo {author} {\bibfnamefont {D.}~\bibnamefont
  {Mehta}}, \bibinfo {author} {\bibfnamefont {N.}~\bibnamefont {Daleo}},
  \bibinfo {author} {\bibfnamefont {F.}~\bibnamefont {D\"orfler}}, \ and\
  \bibinfo {author} {\bibfnamefont {J.~D.}\ \bibnamefont {Hauenstein}},\
  }\href@noop {} {\bibfield  {journal} {\bibinfo  {journal} {Chaos}\ }\textbf
  {\bibinfo {volume} {25}} (\bibinfo {year} {2015})}\BibitemShut {NoStop}%
\bibitem [{\citenamefont {Tinkham}(1975)}]{Tin75}%
  \BibitemOpen
  \bibfield  {author} {\bibinfo {author} {\bibfnamefont {M.}~\bibnamefont
  {Tinkham}},\ }\href@noop {} {\emph {\bibinfo {title} {Introduction to
  Superconductivity}}}\ (\bibinfo  {publisher} {Krieger Publishing Company},\
  \bibinfo {year} {1975})\BibitemShut {NoStop}%
\bibitem [{\citenamefont {Janssens}\ and\ \citenamefont
  {Kamagate}(2003)}]{Jan03}%
  \BibitemOpen
  \bibfield  {author} {\bibinfo {author} {\bibfnamefont {N.}~\bibnamefont
  {Janssens}}\ and\ \bibinfo {author} {\bibfnamefont {A.}~\bibnamefont
  {Kamagate}},\ }\href@noop {} {\bibfield  {journal} {\bibinfo  {journal} {Int.
  J. Elect. Power Energy Syst.}\ }\textbf {\bibinfo {volume} {25}},\ \bibinfo
  {pages} {591 } (\bibinfo {year} {2003})}\BibitemShut {NoStop}%
\bibitem [{\citenamefont {Korsak}(1972)}]{Kor72}%
  \BibitemOpen
  \bibfield  {author} {\bibinfo {author} {\bibfnamefont {A.}~\bibnamefont
  {Korsak}},\ }\href
  {http://ieeexplore.ieee.org/xpls/abs_all.jsp?arnumber=4074824} {\bibfield
  {journal} {\bibinfo  {journal} {IEEE Trans. Power App. Syst.}\ ,\ \bibinfo
  {pages} {1093}} (\bibinfo {year} {1972})}\BibitemShut {NoStop}%
\bibitem [{\citenamefont {Tavora}\ and\ \citenamefont {Smith}(1972)}]{Tav72b}%
  \BibitemOpen
  \bibfield  {author} {\bibinfo {author} {\bibfnamefont {C.~J.}\ \bibnamefont
  {Tavora}}\ and\ \bibinfo {author} {\bibfnamefont {O.~J.}\ \bibnamefont
  {Smith}},\ }\href@noop {} {\bibfield  {journal} {\bibinfo  {journal} {IEEE
  Trans. Power App. Syst.}\ ,\ \bibinfo {pages} {1131}} (\bibinfo {year}
  {1972})}\BibitemShut {NoStop}%
\bibitem [{\citenamefont {Bailleul}\ and\ \citenamefont
  {Byrnes}(1982)}]{Bai82}%
  \BibitemOpen
  \bibfield  {author} {\bibinfo {author} {\bibfnamefont {J.}~\bibnamefont
  {Bailleul}}\ and\ \bibinfo {author} {\bibfnamefont {C.}~\bibnamefont
  {Byrnes}},\ }\href@noop {} {\bibfield  {journal} {\bibinfo  {journal} {Proc.
  of the 21$^{\rm st}$ IEEE Conf. on Decision and Control}\ }\textbf {\bibinfo
  {volume} {2}},\ \bibinfo {pages} {919} (\bibinfo {year} {1982})}\BibitemShut
  {NoStop}%
\bibitem [{\citenamefont {Skar}(1980)}]{Ska80}%
  \BibitemOpen
  \bibfield  {author} {\bibinfo {author} {\bibfnamefont {S.}~\bibnamefont
  {Skar}},\ }\emph {\bibinfo {title} {Stability of Power Systems and other
  Systems of Second Order Differential Equations}},\ \href@noop {} {Ph.D.
  thesis},\ \bibinfo  {school} {Iowa State University} (\bibinfo {year}
  {1980})\BibitemShut {NoStop}%
\bibitem [{\citenamefont {Coletta}\ and\ \citenamefont
  {Jacquod}(2016)}]{Col16}%
  \BibitemOpen
  \bibfield  {author} {\bibinfo {author} {\bibfnamefont {T.}~\bibnamefont
  {Coletta}}\ and\ \bibinfo {author} {\bibfnamefont {P.}~\bibnamefont
  {Jacquod}},\ }\href@noop {} {\bibfield  {journal} {\bibinfo  {journal} {Phys.
  Rev. E}\ }\textbf {\bibinfo {volume} {93}},\ \bibinfo {pages} {032222}
  (\bibinfo {year} {2016})}\BibitemShut {NoStop}%
\bibitem [{\citenamefont {Manik}\ \emph {et~al.}(2014)\citenamefont {Manik},
  \citenamefont {Witthaut}, \citenamefont {Sch\"{a}fer}, \citenamefont
  {Matthiae}, \citenamefont {Sorge}, \citenamefont {Rohden}, \citenamefont
  {Katifori},\ and\ \citenamefont {Timme}}]{Man14}%
  \BibitemOpen
  \bibfield  {author} {\bibinfo {author} {\bibfnamefont {D.}~\bibnamefont
  {Manik}}, \bibinfo {author} {\bibfnamefont {D.}~\bibnamefont {Witthaut}},
  \bibinfo {author} {\bibfnamefont {B.}~\bibnamefont {Sch\"{a}fer}}, \bibinfo
  {author} {\bibfnamefont {M.}~\bibnamefont {Matthiae}}, \bibinfo {author}
  {\bibfnamefont {A.}~\bibnamefont {Sorge}}, \bibinfo {author} {\bibfnamefont
  {M.}~\bibnamefont {Rohden}}, \bibinfo {author} {\bibfnamefont
  {E.}~\bibnamefont {Katifori}}, \ and\ \bibinfo {author} {\bibfnamefont
  {M.}~\bibnamefont {Timme}},\ }\href {\doibase 10.1140/epjst/e2014-02274-y}
  {\bibfield  {journal} {\bibinfo  {journal} {Eur. Phys. J. Special Topics}\
  }\textbf {\bibinfo {volume} {223}},\ \bibinfo {pages} {2527} (\bibinfo {year}
  {2014})}\BibitemShut {NoStop}%
\bibitem [{\citenamefont {Bergen}\ and\ \citenamefont {Hill}(1981)}]{Ber81}%
  \BibitemOpen
  \bibfield  {author} {\bibinfo {author} {\bibfnamefont {A.~R.}\ \bibnamefont
  {Bergen}}\ and\ \bibinfo {author} {\bibfnamefont {D.~J.}\ \bibnamefont
  {Hill}},\ }\href@noop {} {\bibfield  {journal} {\bibinfo  {journal} {IEEE
  Trans. Power App. Syst.}\ }\textbf {\bibinfo {volume} {PAS-100}},\ \bibinfo
  {pages} {25} (\bibinfo {year} {1981})}\BibitemShut {NoStop}%
\bibitem [{\citenamefont {Pai}(1989)}]{Pai89}%
  \BibitemOpen
  \bibfield  {author} {\bibinfo {author} {\bibfnamefont {M.~A.}\ \bibnamefont
  {Pai}},\ }\href@noop {} {\emph {\bibinfo {title} {Energy Function Analysis
  for Power System Stability}}}\ (\bibinfo  {publisher} {Kluwer Academic
  Publishers},\ \bibinfo {year} {1989})\BibitemShut {NoStop}%
\bibitem [{\citenamefont {Backhaus}\ and\ \citenamefont
  {Chertkov}(2013)}]{Bac13}%
  \BibitemOpen
  \bibfield  {author} {\bibinfo {author} {\bibfnamefont {S.}~\bibnamefont
  {Backhaus}}\ and\ \bibinfo {author} {\bibfnamefont {M.}~\bibnamefont
  {Chertkov}},\ }\href {\doibase 10.1063/PT.3.1979} {\bibfield  {journal}
  {\bibinfo  {journal} {Physics Today}\ }\textbf {\bibinfo {volume} {66}},\
  \bibinfo {pages} {42} (\bibinfo {year} {2013})}\BibitemShut {NoStop}%
\bibitem [{\citenamefont {Pecora}\ and\ \citenamefont {Carroll}(1998)}]{Pec98}%
  \BibitemOpen
  \bibfield  {author} {\bibinfo {author} {\bibfnamefont {L.~M.}\ \bibnamefont
  {Pecora}}\ and\ \bibinfo {author} {\bibfnamefont {T.~L.}\ \bibnamefont
  {Carroll}},\ }\href@noop {} {\bibfield  {journal} {\bibinfo  {journal} {Phys.
  Rev. Lett.}\ }\textbf {\bibinfo {volume} {80}},\ \bibinfo {pages} {2109}
  (\bibinfo {year} {1998})}\BibitemShut {NoStop}%
\bibitem [{\citenamefont {Matveev}\ \emph {et~al.}(2002)\citenamefont
  {Matveev}, \citenamefont {Larkin},\ and\ \citenamefont {Glazman}}]{Mat02}%
  \BibitemOpen
  \bibfield  {author} {\bibinfo {author} {\bibfnamefont {K.~A.}\ \bibnamefont
  {Matveev}}, \bibinfo {author} {\bibfnamefont {A.~I.}\ \bibnamefont {Larkin}},
  \ and\ \bibinfo {author} {\bibfnamefont {L.~I.}\ \bibnamefont {Glazman}},\
  }\href@noop {} {\bibfield  {journal} {\bibinfo  {journal} {Phys. Rev. Lett.}\
  }\textbf {\bibinfo {volume} {89}},\ \bibinfo {pages} {096802} (\bibinfo
  {year} {2002})}\BibitemShut {NoStop}%
\bibitem [{\citenamefont {Wiley}\ \emph {et~al.}(2006)\citenamefont {Wiley},
  \citenamefont {Strogatz},\ and\ \citenamefont {Girvan}}]{Wil06}%
  \BibitemOpen
  \bibfield  {author} {\bibinfo {author} {\bibfnamefont {D.~A.}\ \bibnamefont
  {Wiley}}, \bibinfo {author} {\bibfnamefont {S.~H.}\ \bibnamefont {Strogatz}},
  \ and\ \bibinfo {author} {\bibfnamefont {M.}~\bibnamefont {Girvan}},\ }\href
  {\doibase 10.1063/1.2165594} {\bibfield  {journal} {\bibinfo  {journal}
  {Chaos}\ }\textbf {\bibinfo {volume} {16}},\ \bibinfo {pages} {015103}
  (\bibinfo {year} {2006})}\BibitemShut {NoStop}%
\bibitem [{\citenamefont {Menck}\ \emph {et~al.}(2013)\citenamefont {Menck},
  \citenamefont {Heitzig}, \citenamefont {Marwan},\ and\ \citenamefont
  {Kurths}}]{Men13}%
  \BibitemOpen
  \bibfield  {author} {\bibinfo {author} {\bibfnamefont {P.~J.}\ \bibnamefont
  {Menck}}, \bibinfo {author} {\bibfnamefont {J.}~\bibnamefont {Heitzig}},
  \bibinfo {author} {\bibfnamefont {N.}~\bibnamefont {Marwan}}, \ and\ \bibinfo
  {author} {\bibfnamefont {J.}~\bibnamefont {Kurths}},\ }\href@noop {}
  {\bibfield  {journal} {\bibinfo  {journal} {Nat. Phys.}\ }\textbf {\bibinfo
  {volume} {9}},\ \bibinfo {pages} {89} (\bibinfo {year} {2013})}\BibitemShut
  {NoStop}%
\bibitem [{\citenamefont {van Hemmen}\ and\ \citenamefont
  {Wreskinski}(1993)}]{Hem93}%
  \BibitemOpen
  \bibfield  {author} {\bibinfo {author} {\bibfnamefont {J.~L.}\ \bibnamefont
  {van Hemmen}}\ and\ \bibinfo {author} {\bibfnamefont {W.~F.}\ \bibnamefont
  {Wreskinski}},\ }\href@noop {} {\bibfield  {journal} {\bibinfo  {journal} {J.
  Stat. Phys.}\ }\textbf {\bibinfo {volume} {72}},\ \bibinfo {pages} {145}
  (\bibinfo {year} {1993})}\BibitemShut {NoStop}%
\bibitem [{\citenamefont {Araposthatis}\ \emph {et~al.}(1981)\citenamefont
  {Araposthatis}, \citenamefont {Sastry},\ and\ \citenamefont
  {Varayia}}]{Ara81}%
  \BibitemOpen
  \bibfield  {author} {\bibinfo {author} {\bibfnamefont {A.}~\bibnamefont
  {Araposthatis}}, \bibinfo {author} {\bibfnamefont {S.}~\bibnamefont
  {Sastry}}, \ and\ \bibinfo {author} {\bibfnamefont {P.}~\bibnamefont
  {Varayia}},\ }\href@noop {} {\bibfield  {journal} {\bibinfo  {journal} {Int.
  J. Elect. Power Energy Syst.}\ }\textbf {\bibinfo {volume} {3}},\ \bibinfo
  {pages} {115} (\bibinfo {year} {1981})}\BibitemShut {NoStop}%
\bibitem [{\citenamefont {Menck}\ \emph {et~al.}(2014)\citenamefont {Menck},
  \citenamefont {Heitzig}, \citenamefont {Kurths},\ and\ \citenamefont
  {Schellnhuber}}]{Men14}%
  \BibitemOpen
  \bibfield  {author} {\bibinfo {author} {\bibfnamefont {P.~J.}\ \bibnamefont
  {Menck}}, \bibinfo {author} {\bibfnamefont {J.}~\bibnamefont {Heitzig}},
  \bibinfo {author} {\bibfnamefont {J.}~\bibnamefont {Kurths}}, \ and\ \bibinfo
  {author} {\bibfnamefont {H.~J.}\ \bibnamefont {Schellnhuber}},\ }\href@noop
  {} {\bibfield  {journal} {\bibinfo  {journal} {Nat. Comms.}\ }\textbf
  {\bibinfo {volume} {5}},\ \bibinfo {pages} {3969} (\bibinfo {year}
  {2014})}\BibitemShut {NoStop}%
\bibitem [{\citenamefont {Witthaut}\ and\ \citenamefont {Timme}(2012)}]{Wit12}%
  \BibitemOpen
  \bibfield  {author} {\bibinfo {author} {\bibfnamefont {D.}~\bibnamefont
  {Witthaut}}\ and\ \bibinfo {author} {\bibfnamefont {M.}~\bibnamefont
  {Timme}},\ }\href@noop {} {\bibfield  {journal} {\bibinfo  {journal} {New
  Journal of Physics}\ }\textbf {\bibinfo {volume} {14}},\ \bibinfo {pages}
  {083036} (\bibinfo {year} {2012})}\BibitemShut {NoStop}%
\bibitem [{\citenamefont {Lehmann}\ and\ \citenamefont
  {Bernasconi}(2010)}]{Leh10}%
  \BibitemOpen
  \bibfield  {author} {\bibinfo {author} {\bibfnamefont {J.}~\bibnamefont
  {Lehmann}}\ and\ \bibinfo {author} {\bibfnamefont {J.}~\bibnamefont
  {Bernasconi}},\ }\href@noop {} {\bibfield  {journal} {\bibinfo  {journal}
  {Phys. Rev. E}\ }\textbf {\bibinfo {volume} {81}},\ \bibinfo {pages} {031129}
  (\bibinfo {year} {2010})}\BibitemShut {NoStop}%
\bibitem [{\citenamefont {Vaiman}\ \emph {et~al.}(2012)\citenamefont {Vaiman},
  \citenamefont {Bell}, \citenamefont {Chen}, \citenamefont {Chowdhury},
  \citenamefont {Dobson}, \citenamefont {Hines}, \citenamefont {Papic},
  \citenamefont {Miller},\ and\ \citenamefont {Zhang}}]{Vai12}%
  \BibitemOpen
  \bibfield  {author} {\bibinfo {author} {\bibfnamefont {M.}~\bibnamefont
  {Vaiman}}, \bibinfo {author} {\bibfnamefont {K.}~\bibnamefont {Bell}},
  \bibinfo {author} {\bibfnamefont {Y.}~\bibnamefont {Chen}}, \bibinfo {author}
  {\bibfnamefont {B.}~\bibnamefont {Chowdhury}}, \bibinfo {author}
  {\bibfnamefont {I.}~\bibnamefont {Dobson}}, \bibinfo {author} {\bibfnamefont
  {P.}~\bibnamefont {Hines}}, \bibinfo {author} {\bibfnamefont
  {M.}~\bibnamefont {Papic}}, \bibinfo {author} {\bibfnamefont
  {S.}~\bibnamefont {Miller}}, \ and\ \bibinfo {author} {\bibfnamefont
  {P.}~\bibnamefont {Zhang}},\ }\href@noop {} {\bibfield  {journal} {\bibinfo
  {journal} {IEEE Transactions Power Systems}\ }\textbf {\bibinfo {volume}
  {27}},\ \bibinfo {pages} {631} (\bibinfo {year} {2012})}\BibitemShut
  {NoStop}%
\bibitem [{\citenamefont {Pahwa}\ \emph {et~al.}(2014)\citenamefont {Pahwa},
  \citenamefont {Scoglio},\ and\ \citenamefont {Scala}}]{Pah14}%
  \BibitemOpen
  \bibfield  {author} {\bibinfo {author} {\bibfnamefont {S.}~\bibnamefont
  {Pahwa}}, \bibinfo {author} {\bibfnamefont {C.}~\bibnamefont {Scoglio}}, \
  and\ \bibinfo {author} {\bibfnamefont {A.}~\bibnamefont {Scala}},\
  }\href@noop {} {\bibfield  {journal} {\bibinfo  {journal} {Sci. Rep.}\
  }\textbf {\bibinfo {volume} {4}},\ \bibinfo {pages} {3694} (\bibinfo {year}
  {2014})}\BibitemShut {NoStop}%
\bibitem [{\citenamefont {Kuramoto}(1984)}]{Kur84}%
  \BibitemOpen
  \bibfield  {author} {\bibinfo {author} {\bibfnamefont {Y.}~\bibnamefont
  {Kuramoto}},\ }\href@noop {} {\bibfield  {journal} {\bibinfo  {journal}
  {Progr. Theoret. Phys. Suppl.}\ }\textbf {\bibinfo {volume} {79}},\ \bibinfo
  {pages} {223} (\bibinfo {year} {1984})}\BibitemShut {NoStop}%
\bibitem [{\citenamefont {Strogatz}(2000)}]{Str00}%
  \BibitemOpen
  \bibfield  {author} {\bibinfo {author} {\bibfnamefont {S.}~\bibnamefont
  {Strogatz}},\ }\href@noop {} {\bibfield  {journal} {\bibinfo  {journal}
  {Physica D}\ }\textbf {\bibinfo {volume} {143}},\ \bibinfo {pages} {1}
  (\bibinfo {year} {2000})}\BibitemShut {NoStop}%
\bibitem [{\citenamefont {Acebr\'on}\ \emph {et~al.}(2005)\citenamefont
  {Acebr\'on}, \citenamefont {Bonilla}, \citenamefont {P\'erez~Vicente},
  \citenamefont {Ritort},\ and\ \citenamefont {Spigler}}]{Ace05}%
  \BibitemOpen
  \bibfield  {author} {\bibinfo {author} {\bibfnamefont {J.}~\bibnamefont
  {Acebr\'on}}, \bibinfo {author} {\bibfnamefont {L.}~\bibnamefont {Bonilla}},
  \bibinfo {author} {\bibfnamefont {C.}~\bibnamefont {P\'erez~Vicente}},
  \bibinfo {author} {\bibfnamefont {F.}~\bibnamefont {Ritort}}, \ and\ \bibinfo
  {author} {\bibfnamefont {R.}~\bibnamefont {Spigler}},\ }\href@noop {}
  {\bibfield  {journal} {\bibinfo  {journal} {Rev. Mod. Phys.}\ }\textbf
  {\bibinfo {volume} {77}},\ \bibinfo {pages} {137} (\bibinfo {year}
  {2005})}\BibitemShut {NoStop}%
\bibitem [{\citenamefont {Arenas}\ \emph {et~al.}(2008)\citenamefont {Arenas},
  \citenamefont {D{\'\i}az-Guilera}, \citenamefont {Kurths}, \citenamefont
  {Moreno},\ and\ \citenamefont {Zhou}}]{Are08}%
  \BibitemOpen
  \bibfield  {author} {\bibinfo {author} {\bibfnamefont {A.}~\bibnamefont
  {Arenas}}, \bibinfo {author} {\bibfnamefont {A.}~\bibnamefont
  {D{\'\i}az-Guilera}}, \bibinfo {author} {\bibfnamefont {J.}~\bibnamefont
  {Kurths}}, \bibinfo {author} {\bibfnamefont {Y.}~\bibnamefont {Moreno}}, \
  and\ \bibinfo {author} {\bibfnamefont {C.}~\bibnamefont {Zhou}},\ }\href@noop
  {} {\bibfield  {journal} {\bibinfo  {journal} {Physics Reports}\ }\textbf
  {\bibinfo {volume} {469}},\ \bibinfo {pages} {93} (\bibinfo {year}
  {2008})}\BibitemShut {NoStop}%
\bibitem [{\citenamefont {Wiesenfeld}\ \emph {et~al.}(1998)\citenamefont
  {Wiesenfeld}, \citenamefont {Colet},\ and\ \citenamefont {Strogatz}}]{Wie98}%
  \BibitemOpen
  \bibfield  {author} {\bibinfo {author} {\bibfnamefont {K.}~\bibnamefont
  {Wiesenfeld}}, \bibinfo {author} {\bibfnamefont {P.}~\bibnamefont {Colet}}, \
  and\ \bibinfo {author} {\bibfnamefont {S.}~\bibnamefont {Strogatz}},\
  }\href@noop {} {\bibfield  {journal} {\bibinfo  {journal} {Phys. Rev. E}\
  }\textbf {\bibinfo {volume} {57}},\ \bibinfo {pages} {1563} (\bibinfo {year}
  {1998})}\BibitemShut {NoStop}%
\bibitem [{\citenamefont {Rogge}\ and\ \citenamefont {Aeyels}(2004)}]{Rog04}%
  \BibitemOpen
  \bibfield  {author} {\bibinfo {author} {\bibfnamefont {J.~A.}\ \bibnamefont
  {Rogge}}\ and\ \bibinfo {author} {\bibfnamefont {D.}~\bibnamefont {Aeyels}},\
  }\href@noop {} {\bibfield  {journal} {\bibinfo  {journal} {J. Phys. A}\
  }\textbf {\bibinfo {volume} {37}},\ \bibinfo {pages} {11135} (\bibinfo {year}
  {2004})}\BibitemShut {NoStop}%
\bibitem [{\citenamefont {Ochab}\ and\ \citenamefont {G\'ora}(2010)}]{Och10}%
  \BibitemOpen
  \bibfield  {author} {\bibinfo {author} {\bibfnamefont {J.}~\bibnamefont
  {Ochab}}\ and\ \bibinfo {author} {\bibfnamefont {P.~F.}\ \bibnamefont
  {G\'ora}},\ }\href@noop {} {\bibfield  {journal} {\bibinfo  {journal} {Acta
  Phys. Pol. B [Proc. Suppl. 3]}\ }\textbf {\bibinfo {volume} {3}},\ \bibinfo
  {pages} {453} (\bibinfo {year} {2010})}\BibitemShut {NoStop}%
\bibitem [{\citenamefont {Heitmann}\ and\ \citenamefont
  {Ermentrout}(2015)}]{Hei15}%
  \BibitemOpen
  \bibfield  {author} {\bibinfo {author} {\bibfnamefont {S.}~\bibnamefont
  {Heitmann}}\ and\ \bibinfo {author} {\bibfnamefont {G.~B.}\ \bibnamefont
  {Ermentrout}},\ }\href@noop {} {\bibfield  {journal} {\bibinfo  {journal}
  {Biol. Cybern.}\ }\textbf {\bibinfo {volume} {109}},\ \bibinfo {pages} {333}
  (\bibinfo {year} {2015})}\BibitemShut {NoStop}%
\bibitem [{\citenamefont {Filatrella}\ \emph {et~al.}(2008)\citenamefont
  {Filatrella}, \citenamefont {Nielsen},\ and\ \citenamefont
  {Pedersen}}]{Fil08}%
  \BibitemOpen
  \bibfield  {author} {\bibinfo {author} {\bibfnamefont {G.}~\bibnamefont
  {Filatrella}}, \bibinfo {author} {\bibfnamefont {A.~H.}\ \bibnamefont
  {Nielsen}}, \ and\ \bibinfo {author} {\bibfnamefont {N.~F.}\ \bibnamefont
  {Pedersen}},\ }\href@noop {} {\bibfield  {journal} {\bibinfo  {journal} {The
  European Physical Journal B}\ }\textbf {\bibinfo {volume} {61}},\ \bibinfo
  {pages} {485} (\bibinfo {year} {2008})}\BibitemShut {NoStop}%
\bibitem [{\citenamefont {Chopra}\ and\ \citenamefont {Spong}(2009)}]{Cho09}%
  \BibitemOpen
  \bibfield  {author} {\bibinfo {author} {\bibfnamefont {N.}~\bibnamefont
  {Chopra}}\ and\ \bibinfo {author} {\bibfnamefont {M.~W.}\ \bibnamefont
  {Spong}},\ }\href {\doibase 10.1109/TAC.2008.2007884} {\bibfield  {journal}
  {\bibinfo  {journal} {IEEE Transactions on Automatic Control}\ }\textbf
  {\bibinfo {volume} {54}},\ \bibinfo {pages} {353} (\bibinfo {year}
  {2009})}\BibitemShut {NoStop}%
\bibitem [{\citenamefont {Biggs}(1993)}]{Big93}%
  \BibitemOpen
  \bibfield  {author} {\bibinfo {author} {\bibfnamefont {N.}~\bibnamefont
  {Biggs}},\ }\href@noop {} {\emph {\bibinfo {title} {Algebraic graph
  theory}}},\ \bibinfo {edition} {2nd}\ ed.\ (\bibinfo  {publisher} {Cambridge
  University Press},\ \bibinfo {year} {1993})\BibitemShut {NoStop}%
\end{thebibliography}%
\end{document}